\RequirePackage{ifpdf}
\documentclass[11pt, a4paper]{article}
\pdfoutput=1
\usepackage{jheppub}
\usepackage{epsfig}
\usepackage[utf8]{inputenc}
\usepackage{graphicx}
\usepackage{amssymb,amsmath}
\usepackage{bbold}
\usepackage{fancybox,framed}
\usepackage{dsfont}
\usepackage{braket}
\usepackage{slashed}
\usepackage{comment}
\usepackage{xcolor}
\usepackage{soul}

\usepackage{caption}
\usepackage{subcaption}

\csname @addtoreset\endcsname{equation}{section}

\def\bseq{\begin{subequation}}  
\def\eseq{\end{subequation}}
\def\bsea{\begin{subeqnarray}}  
\def\esea{\end{subeqnarray}}

\newcommand{\beq}{\begin{equation}}
\newcommand{\bea}{\begin{eqnarray}}
\newcommand{\eea}{\end{eqnarray}}
\newcommand{\eeq}{\end{equation}}

\newcommand{\mc}{\mathcal}

\renewcommand{\d}{\delta}
\newcommand{\pa}{\partial}
\newcommand{\del}{\partial}

\newcommand{\m}{\mu}

\newcommand{\n}{\nu}

\newcommand{\p}{\pi}

\renewcommand{\r}{\rho}

\renewcommand{\S}{\Sigma}

\renewcommand{\O}{\Omega}

\newcommand{\mt}[1]{\textrm{\tiny #1}}

\newcommand{\reef}[1]{(\ref{#1})}

\newcommand{\eg}{{\it e.g.,}\ }
\newcommand{\ie}{{\it i.e.,}\ }
\def\({\left(} \def\){\right)}
\def\[{\left[} \def\]{\right]}
\def\del{{\partial}}

\newcommand{\see}{S_\mt{EE}}
\newcommand{\tr}{\textup{Tr}}

\newcommand{\be}{\beta}

\newcommand{\de}{\delta}

\newcommand{\la}{\lambda}

\renewcommand{\r}{\rho}

\newcommand{\Si}{\Sigma}

\renewcommand{\O}{\Omega}

\newcommand{\ren}{R\'enyi}
\newcommand{\te}{t_\mt{E}}
\newcommand{\conj}{\text{conj}}

\declareslashed{}{\backslash}{0}{0}{\Omega}
\declareslashed{}{\backslash}{0}{0}{O}

\def\vev#1{\langle #1 \rangle}
\newcommand{\lp}{\ell_{\mt P}}

\newcommand{\Tr}{\text{Tr}}

\newcommand{\rev}[1]{#1}

\author[a]{Lorenzo Bianchi,}
\author[b]{Shira Chapman,}
\author[c]{Xi Dong,}
\author[b,d]{Dami\'an A. Galante,}
\author[b,e]{Marco Meineri}
\author[b]{and Robert C. Myers}

\affiliation[a]{Institut f\"ur Theoretische Physik,
Universit\"at Hamburg\\
Luruper Chaussee 149,
22761 Hamburg, Germany}

\affiliation[b]{Perimeter Institute for Theoretical Physics\\
31 Caroline Street North, ON N2L 2Y5, Canada}

\affiliation[c]{School of Natural Sciences, Institute for Advanced Study\\
1 Einstein Drive, Princeton, New Jersey 08540, USA}

\affiliation[d]{Department of Applied Mathematics,
University of Western Ontario,\\
London, Ontario N6A 5B7, Canada}

\affiliation[e]{Scuola Normale Superiore, Piazza dei Cavalieri 7 I-56126 Pisa, Italy\\
and Istituto Nazionale di Fisica Nucleare - sezione di Pisa}

\emailAdd{lorenzo.bianchi@desy.de}
\emailAdd{schapman@perimeterinstitute.ca}
\emailAdd{xidong@ias.edu}
\emailAdd{dgalante@perimeterinstitute.ca}
\emailAdd{marco.meineri@sns.it}
\emailAdd{rmyers@perimeterinstitute.ca}

\abstract{We present a holographic method for computing the response of \ren\ entropies in conformal field theories to small shape deformations around a flat (or spherical) entangling surface.
Our strategy employs the stress tensor one-point function in a deformed hyperboloid background and relates it to
the coefficient in the two-point function of the displacement operator.
We obtain explicit numerical results for $d=3,\cdots, 6$ spacetime dimensions, and also evaluate analytically the limits where the \ren\ index approaches 1 and 0 in general dimensions. We use our results to extend the work of 1602.08493 and disprove a set of conjectures in the literature regarding the relations between the \ren\ shape dependence and the conformal weight of the twist operator. We also extend our analysis beyond leading order in derivatives in the bulk theory by studying Gauss-Bonnet gravity.
}

\title{Shape Dependence of Holographic \ren\ Entropy in General Dimensions}

\keywords{\ren\ entropy, shape deformations, displacement operator, conformal defect}

\begin{document}

\maketitle

\section{Introduction and Summary}
\label{intro}

Entanglement is one of the key features which distinguishes quantum physics from the classical realm and it is widely recognized as an essential ingredient in shaping many of the physical properties of complex interacting quantum systems. In particular, there is an increasing realization of the important role which entanglement plays in quantum field theory (QFT) \cite{Calabrese:2004eu,Calabrese:2005in,Calabrese:2005zw} and quantum gravity \cite{qg1,qg2,qg3,qg4}. While there are a variety of measures of entanglement \cite{qibook}, two which have received particular attention in the latter fields are entanglement and \ren\ entropies.
For example, typical calculations begin with some QFT in a (global) state described by the density matrix $\r$  on a given time slice. Then one restricts the state to a particular region $A$ by tracing over the degrees of freedom in the complementary region $B$ to produce:
\begin{equation}
\r_\mt{A} = tr_B(\r)\,. \label{reduce}
\end{equation}
The above entanglement measures are constructed from the reduced density matrix as
\bea
\see& = &- \tr (\r_\mt{A} \log \r_\mt{A})\, , \label{EEdef}\\
S_n &=& \frac{1}{1-n}\, \log \tr (\r_\mt{A}^n)\,.
\label{renyidef}
\eea
In particular, the \ren\ entropies $S_n$ form a one-parameter family labeled by the index $n$, which is often taken to be an integer (with $n>1$)   \cite{renyi1,renyi2}. However, when $S_n$ can be continued to real $n$, the entanglement entropy can be  recovered with the limit: $S_\mt{EE}=\lim_{n\to 1} S_n$.

From a certain perspective, the \ren\ entropies \reef{renyidef} are `less complicated' objects than the entanglement entropy \reef{EEdef}. One manifestation of this assertion is that $S_n$ can be evaluated as the expectation value of an operator in (a replicated version of) the QFT for integer $n>1$. In particular, eq.~\reef{renyidef}
can be recast as
\beq
S_n=\frac{1}{1-n}\, \log \braket{\tau_n}\,,
\label{renyidef2}
\eeq
where the twist operator $\tau_n$ is a codimension-two surface operator with support on the entangling surface (which divides the time slice into regions $A$ and $B$) \cite{Calabrese:2004eu,Cardy:2007mb,Hung:2014npa}. To be precise, the above expectation value is taken in the tensor product of $n$ copies of the QFT
-- see further details in section \ref{cftside}. This reformulation of the \ren\ entropies also allows these quantities to be evaluated using quantum Monte Carlo techniques, \eg \cite{roger1,roger2,roger3}, and even to be measured in the laboratory, \eg \cite{dima,expEE}.

However, turning to holography, the situation is somehow reversed. In the context of the AdS/CFT correspondence, the RT and HRT prescriptions \cite{Ryu:2006bv,Ryu:2006ef,Hubeny:2007xt} provide an elegant geometric tool which can be implemented in a straightforward fashion to evaluate the entanglement entropy in the boundary theory for general situations.\footnote{This approach has also been extended to include higher curvature interactions in the gravitational dual \cite{Hung:2011xb,deBoer:2011wk,Dong:2013qoa,Camps:2013zua}, as well as quantum fluctuations in the bulk \cite{Faulkner:2013ana}.} The recent derivations \cite{Lewkowycz:2013nqa,DLR} of these two prescriptions also yield a geometric construction to evaluate holographic \ren\ entropies, which can be formulated as evaluating the area of a cosmic brane in a backreacted bulk geometry \cite{Dong:2016fnf}. Unfortunately, this approach does not yield a practical calculation except in very special situations. One example is the case of a spherical entangling surface in a boundary conformal field theory (CFT)  \cite{Casini:2011kv, Hung:2011nu} where the backreacted geometry becomes a hyperbolic black hole in AdS, as will be reviewed in section \ref{hollow}. Further, progress in this direction was made recently \cite{Dong:2016wcf} by studying the variations of $S_n$ for small perturbations of a spherical entangling surface for a four-dimensional boundary CFT -- see also \cite{Camps:2016gfs}. In this paper, we provide a generalization of these calculations \cite{Dong:2016wcf} to any number of boundary dimensions.

Our investigation relies on the field theoretic approach introduced in \cite{Bianchi:2015liz} to investigating the shape dependence of \ren\ entropies in CFTs.\footnote{\rev{For previous studies of the shape dependence of entanglement entropy, see \cite{mark0,Bueno:2015rda,Bueno:2015lza,Carmi:2015dla,Fonda:2014cca,Fonda:2015nma}.}} In particular, they examined the twist operators as conformal defects, \eg \cite{Cardy:1984bb,McAvity:1995zd,Billo:2016cpy}. This framework naturally leads to the definition of the displacement operator, which implements small local deformations of the entangling surface. Further, this work allowed a variety of different conjectures, \rev{\eg \cite{safdi1,Lewkowycz:2014jia,Bueno:2015qya,Bueno:2015lza,Bianchi:2015liz}} with regards to the shape dependence of $S_n$ to be consolidated in terms of a single simple constraint \cite{Bianchi:2015liz}:
\beq
C_D(n)= d\, \Gamma\!\left(\tfrac{d+1}{2}\right)\, \left(\tfrac{2}{\sqrt{\pi}}\right)^{d-1}\, h_n\,,
\label{conj}
\eeq
for a $d$-dimensional CFT. Here, $C_D(n)$ is the coefficient defining the two-point function of the displacement operator and $h_n$ is the conformal weight of the twist operator, which controls the correlator of the stress tensor with the twist operator -- see eq.~\reef{Tflat}. This constraint is known to hold for free massless scalars and fermions in $d=3$ \cite{Dowker1,Dowker2}, as well as free massless scalars in $d=4$ \cite{Bianchi:2015liz}. Further, the $n\to 1$ limit of eq.~\reef{conj} was recently proven to hold in general CFTs  \cite{Faulkner:2015csl}.\footnote{Of course, both $C_D(n)$ and $h_n$ vanish at $n=1$. Hence the nontrivial result of \cite{Faulkner:2015csl} is that the first derivative of eq.~\reef{conj} with respect to $n$ holds at $n=1$.} However, eq.~\reef{conj} is {\it not} a universal relation for general CFTs at general values of $n$. In particular,  the results of \cite{Dong:2016wcf} imply that this  constraint fails for four-dimensional holographic CFTs. With our extension of these holographic calculations to general dimensions, we will explicitly confirm that eq.~\reef{conj} does not hold for holographic CFTs in any dimension.

The paper is organized as follows: In section \ref{cftside}, we review in detail the defect CFT language, and we show how it allows one to generalize the results of \cite{Dong:2016wcf} to arbitrary number of dimensions. In particular, we show that $C_D$ appears in the expectation value of the stress tensor in the presence of a deformed defect (entangling surface). In section \ref{hollow}, we review the construction of the holographic dual of a deformed planar entangling surface, and
the determination of $C_D$ by simply extracting the expectation value of the stress tensor in this background. In section \ref{Einstein}, we perform this computation numerically in the holographic dual of Einstein gravity in $3\leq d \leq 6$,  as well as in an analytic expansion  around $n=1$ to order $(n-1)^2$ in any number of dimensions. We also examine the limit $n\to0$ for general dimensions, which is amenable to analytic result. In section \ref{gaussbonnet}, we then probe the dependence of $C_D$ on higher derivative corrections in the bulk, by adding a Gauss-Bonnet curvature-squared interaction. We extract $C_D$ numerically in $d=4,\,5$, as well as to second order in an analytic expansion around $n=1$ in $4\leq d \leq 6$. From the latter, we observe that for a special value of the Gauss-Bonnet coupling, eq.~\eqref{conj} holds to order $(n-1)^2$. Finally, we obtain analytically the value of $C_D$ in the limit $n\to0$ for any number of dimensions, and find that the result is independent of the Gauss-Bonnet coupling and hence, matches the corresponding result for Einstein gravity. We conclude in section \ref{sec:discussion} with a brief discussion of our results.
Some technical details are relegated to the appendices: Appendix \ref{deltas} provides the details needed to derive a certain useful representation of the two-point correlators used in section \ref{cftside}.
In appendix \ref{app_holo_ren}, we describe how the expression for the boundary stress tensor used in section \ref{holgra} is determined through holographic renormalization with Einstein gravity in the bulk.

Before proceeding, let us finally emphasize that our procedure applies equally well to any other conformal defect: the only place in which the information about \ren\ entropy enters the computation is in the specific form of the dual metric. Finally, let us add that while this paper was in the final stages of preparation, ref.~\cite{again} appeared with results similar to those in section \ref{Einstein}.

\section{The CFT Story}
\label{cftside}

The main object of interest here will be the twist operators which appear in evaluating the \ren\ entropies as in eq.~\reef{renyidef2}. These operators are best understood for two-dimensional CFTs since in this context, they are {\it local} primary operators \cite{Calabrese:2004eu,Cardy:2007mb}. For CFTs or more generally QFTs in higher dimensions, twist operators are formally defined with the replica method, \eg see \cite{Hung:2011nu,Hung:2014npa}. However, in higher dimensions, they become nonlocal surface operators and their properties are less well understood. The replica method begins with a Euclidean path integral representation of the reduced density matrix $\rho_\mt{A}$ where independent boundary  conditions are fixed on the region $A$ as it is approached from above and below in Euclidean time, \ie with $\te\to0^\pm$. To evaluate $\tr (\r_\mt{A}^n)$ in eq.~\reef{renyidef} then, the path integral is extended to a path integral on an $n$-sheeted geometry, where the consecutive sheets are sewn together on cuts running over $A$. The result is often expressed as $\tr (\r_\mt{A}^n)={Z_n}/{(Z_1)^n}$ where $Z_n$ is the partition integral on the full  $n$-sheeted geometry.\footnote{The denominator is introduced here to ensure the correct normalization, \ie $\Tr[\rho_\mt{A}]=1$.} To introduce the twist operator $\tau_n[A]$, this construction is replaced by a path integral over $n$ copies of the underlying QFT on a single copy of the background geometry.  The twist operator is then defined as the codimension-two surface operator extending over the entangling surface, whose expectation value yields
\begin{align}
  \langle\, \tau_n[A]\,\rangle = \frac{Z_n}{(Z_1)^n} 
\,,\label{sloppy}
\end{align}
where the expectation value on the left-hand side is taken in the $n$-fold replicated QFT. Hence eq.~\reef{sloppy} implies that $\tau_n[A]$ opens a branch cut over the region $A$ which connects consecutive copies of the QFT in the replicated theory. Note that to reduce clutter in the following, we will omit the $A$ dependence of the twist operators $\tau_n$.

For the remainder of our discussion, we will consider the case where the underlying field theory is a CFT, which allows us to take advantage of the description of the twist operators as conformal defects \cite{Bianchi:2015liz}. Further we will focus on the special case of a planar entangling surface $\Si$, which will allow us to take advantage of the symmetry of the background geometry.\footnote{Planar and spherical entangling surfaces are conformally equivalent and so the following discussion could equally well be formulated in terms of a spherical entangling surface. With regards to eq.~\reef{conj}, we note that both $C_D(n)$ and $h_n$ control short distance singularities in particular correlators involving the twist operators, \eg see eqs.~\reef{D2p} and \reef{Tflat}, and so these parameters characterize general twist operators, independently of the details of the geometry of the entangling surface.} As discussed above and in the introduction, for integer $n>1$, the computation of $S_n$ is related to the expectation value of a twist operator $\tau_n$,
\beq
S_n=\frac{1}{1-n}\, \log \braket{\tau_n}\,,
\label{Sntau}
\eeq
where the expectation value is taken in the tensor product theory $(\textup{CFT})^n$.
The twist operator breaks translational invariance in the directions orthogonal to $\Si$, and correspondingly the Ward identities of the stress tensor acquire an additional contact term at the location of the defect:\footnote{Let us stress that the Ward identity \eqref{Tward}, as usual, should be interpreted as if both sides were inserted in a correlation function.}
\beq
\pa_\m T^{\m a}_\textup{tot}(x,y)=\de_\Si(x)\, D^a(y).
\label{Tward}
\eeq
Here we split the coordinates of the insertion into orthogonal ($x^a$) and parallel ($y^i$) ones, \ie the defect sits at $x^a=0$. We shall also sometimes regroup them as $z^\m=(x^a,y^i)$. The subscript `tot' in eq.~\eqref{Tward} indicates that the stress tensor is the total stress tensor of the full replicated theory, $(\textup{CFT})^n$ -- equivalently, it is inserted in all the copies of the replicated geometry. In the absence of this subscript, $T_{\m\n}$ refers to the stress tensor of a single copy of the CFT. The delta function $\de_\Si$ has support on the twist operator. Hence, the operator $D^a$, which is known as the \emph{displacement operator}, lives on this defect. If we denote the position of the twist operator in space with $X^\m(y)$ -- \ie in the present (planar) case, $X^\m=(0,y^i)$ -- and the unit vectors orthogonal to the defect with $n^\m_a$, we can give a definition of the displacement in terms of the correlator $\braket{ \cdots\,}_n$ of arbitrary insertions in the presence of the twist operator
\beq
\braket{D^a \cdots\,}_n=n^a_\m\,\frac{\de}{\de X_\m} \braket{ \cdots\,}_n.
\label{Ddef}
\eeq
In the above expression and throughout the following,  expectation values labeled by $n$ are implicitly taken in the presence of the twist operator. Furthermore, recall that in the present discussion, $\tau_n$ has support on the flat entangling surface $\Si$ -- in a more general case, eq.~\eqref{Ddef} would compute the connected part of the correlator. The definition \eqref{Ddef}
makes it obvious that, much like a diffeomorphism is equivalent to the insertion of $\de g_{\m\n} T^{\m\n}$ in the path integral, the response of a defect to a displacement
\begin{equation}
\de X^\m=\de^\m_a f^a,
\end{equation} is given by repeated insertions of the displacement operator, \eg
\beq
(1-n)\,\de S_n = \frac{1}{2}\int_\Si\!dw \int_\Si\!dw'\, f^a(w)f^b(w') \braket{D_a(w)D_b(w')}_n+O(f^4).
\label{deS}
\eeq
In eq.~\eqref{deS}, we disregarded the insertion of the contributions of a single $D^a$ since the one-point function of $D^a$ vanishes for a flat (or spherical) defect.\footnote{The same holds for the three-point function with a flat entangling surface.} The two-point function of the displacement operator is fixed up to a single coefficient
\beq
\braket{D_a(w)D_b(w')}_n =\de_{ab}\, \frac{C_D}{(w-w')^{2(d-1)}}\,.
\label{D2p}
\eeq

Of course, $C_D$ is the parameter which we wish to determine here. Extracting it from a direct computation of $\de S_n$ in eq.~\eqref{deS} would involve second order perturbation theory around a flat entangling surface. Luckily, $C_D$ appears in other observables, some of which are linear in the displacement operator, and so will require only a leading order perturbation. It is convenient to focus on the correlation function between the displacement operator and the stress tensor. The generic two-point function of primaries (in the presence of a planar defect) with the relevant quantum numbers was given in \cite{Billo:2016cpy} in terms of three OPE coefficients $b_i$
\begin{subequations}
\begin{align}
& \braket{D_a(w)T_{ij}(z)}_n=x_a\left[\frac{(b_2-b_1)\, \d_{ij}}{d\,  x^2 (x^2+w^2)^{d-1}}+ \frac{4 b_1\, w_i w_j}{(x^2+w^2)^{d+1}} \right],
 \\
& \braket{D_a(w)T_{bi}(z)}_n=w_i \left[\frac{b_3 \delta_{a b}} {\left(x^2+w^2\right)^{d}}-\frac{x_a x_b \left(\left(b_3-2 b_1\right) w^2+\left(2 b_1+b_3\right) x^2\right)
   }{x^2\left(x^2+w^2\right)^{d+1}}\right],
   \\
\begin{split}
& \braket{D_a(w)T_{bc}(z)}_n =
\\
&~~~~~~~~~~\frac{b_3}{2x^2\left(x^2+w^2\right)^{d}} \left[-x_a \delta _{b c} \left(x^2+w^2\right)+x_c \delta_{a b}\left(w^2-x^2\right) +x_b\delta_{a c} \left(w^2-x^2\right)  \right]\\
  & ~~~~~ +\frac{x_a x_b x_c}{x^4\left(x^2+w^2\right)^{d+1}} \left[\left(b_1+b_2-b_3\right) w^4+2 \left(b_2-b_1\right) x^2 w^2+\left(b_1+b_2+b_3\right) x^4\right],
\end{split}
\end{align}
\label{TD}%
\end{subequations}
where we recall that $z^\m=(x^a,y^i)$, but we have further fixed $y^i=0$ in these expressions. In fact, when the operators involved are the displacement operator and the stress tensor, only two of the three coefficients are linearly independent:
\begin{subequations}
\begin{align}
b_1&= \frac{(d-1) }{\pi (d-2)} \frac{C_D}{n}-\frac{2^{d-2} \pi ^{-\frac{d+1}{2}} d^2  \Gamma \left(\frac{d+1}{2}\right)}{d-2}\, \frac{h_n}{n} \label{eq:b1}\\
b_2&= -\frac{1}{\pi(d-2) } \frac{C_D}{n}+\frac{2^{d-2} \pi ^{-\frac{d+1}{2}} d^2 \Gamma \left(\frac{d+1}{2}\right) }{d-2}\, \frac{h_n}{n}\\
b_3&= 2^{d-1} \pi ^{-\frac{d+1}{2}}d\,  \Gamma \left(\frac{d+1}{2}\right) \frac{h_n}{n}.\label{b123}%
\end{align}
\end{subequations}
The coefficient $h_n$ which appears in these expressions is the so-called \emph{conformal weight} of the twist operator. It is defined by the expectation value of the stress tensor with a planar twist operator\footnote{We emphasize again that the stress tensor here acts in single copy of the CFT and hence there is a factor of $1/n$ on the right-hand side, \eg compare to eq.~(2.12) in \cite{Hung:2014npa}.}
\begin{equation}
 \braket{{T}_{ij}(z)}_n =-\frac{h_n}{2\pi n}\frac{\d_{ij}}{|x|^d},\qquad
 \braket{{T}_{ab}(z)}_n =\frac{h_n}{2\p n}\frac{1}{|x|^d}\left((d-1)\, \d_{ab}-d\, \frac{x_a x_b}{x^2}\right).
 \label{Tflat}
\end{equation}
Now we can use eq.~\eqref{Ddef} to compute the same expectation value, but in the presence of the deformed entangling surface ($f\Si$)
\begin{equation}
 \braket{{T}_{\m\n}(z)}_{n,f\S}=\braket{T_{\m\n}(z)}_n - \int d^{d-2} w \braket{D_a(w)T_{\m\n}(z)}_n f^a(w)+O\!\left(f^2\right).
\label{TfSi}
\end{equation}
Clearly, for a generic deformation the integral cannot be performed. However, it turns out that the singular terms in the short distance expansion $|x|\to0$ can be written down explicitly. This is due to the following property of the correlation function \eqref{TD}. When the limit $|x|\to0$ is taken in the weak sense,  \ie after integration against a test function, the first few coefficients in the expansion are distributions with support at $w=0$. More precisely, the following formula is proven in appendix \ref{deltas}:
\begin{subequations}
\begin{align}
\begin{split}\label{Tdeltas_a_ij}
  \langle D_a(w) T_{ij}(z)\rangle_n=&\frac{x_a}{|x|^d}\left[
  B_1\left(\frac{1}{|x|^2}+\frac{\pa^2 }{2(d-2)} \right)\de^{d-2}(w)\delta_{ij}\right. \\
 & \left.~~~+B_4
  \left(\pa_i \pa_j-\de_{ij}\frac{\pa^2}{d-2}\right)\de^{d-2}(w)\right] + \ldots,
\end{split}
\\
\begin{split}\label{Tdeltas_a_bi}
 \langle D_a(w) T_{bi}(z) \rangle_n=&
-\frac{\pa_i \de^{d-2}(w)}{\left|x|^d\right.}
B_1\left(\delta_{ab}-\frac{x_ax_b}{x^2}\right)
\\
&-\frac{\pa_i\pa^2 \de^{d-2}(w)}{(d-2)\, |x|^{d-2}}
  \left(
 \frac{B_1}{2}
  \delta_{ab}  + B_3\frac{x_ax_b}{x^2} \right) + \dots  ,
\end{split}
  \\
\label{Tdeltas_a_bc}%
 \langle D_a(w)T_{bc}(z)\rangle_n=&
  -\frac{\delta^{d-2}(w)}{|x|^{d+2}} \,\, B_1
 \left[2\delta_{a(b} x_{c)}+x_a \left((d-1)  \delta_{bc}-\frac{ x_b x_c}{|x|^2}(d+2)\right)\right]
 \\
  +&
\frac{\pa^2 \de^{d-2}(w)}{2(d-2)|x|^d} B_1 \left[
2 \de_{a(b} x_{c)}- x_a
 \left((d-1)\d_{bc}-\frac{x_bx_c}{|x|^2}(d-2)\right)\right] + \ldots,\nonumber
\end{align}
\end{subequations}
with
 \begin{align}
 B_1&=\frac{d h_n}{2 \pi n}\,,
 &
  B_2&=\frac{(d-1)   \Gamma \left(\frac{d}{2}-1\right)\pi ^{\frac{d}{2}-2}}{2\Gamma (d+1)}\,\frac{C_D}{n}\,, \notag\\
 B_3&= B_2-\frac{d\, B_1}{2(d-2)} \,,
 &
 B_4&= B_2-\frac{B_1}{d-2}\,.
 \end{align}
The ellipsis in eqs.~\eqref{Tdeltas_a_ij}-\eqref{Tdeltas_a_bc} stand for terms which are less singular in the distance from the entangling surface, and which we will not need in this work. These terms can however be fully expressed using the formulas of appendix \ref{deltas}.
The fact that the singular terms are local imply, via eq.~\eqref{TfSi}, that in the limit of short distance from the defect $\braket{{T}_{\m\n}(z)}_{f\S}$ depends locally on derivatives of the deformation $f^a(y)$.
One more comment is in order. The one-point function in eq.~\eqref{Tflat} refers to a flat entangling surface. Of course, the one-point function in the presence of a defect obtained from this one via a conformal transformation is still proportional to $h_n$. Correspondingly, in eqs.~\eqref{Tdeltas_a_ij}-\eqref{Tdeltas_a_bc}, $C_D$ only appears as a coefficient of the traceless part of $\pa_i\pa_j \de(w)$ and of the third derivative $\pa_i\pa^2 \de(w)$. Indeed, recall that at leading order the extrinsic curvature is $K_{ij}^a=-\pa_i\pa_j f^a$, \eg see \cite{Bianchi:2015liz}, and that conformal transformations map planes into spheres, whose extrinsic curvature is proportional to the identity and constant.

\subsection{Adapted Coordinates}\label{adapted}

In view of the holographic computation in the next section, it is useful to write the one-point function \eqref{TfSi} in a coordinate system adapted to the shape of the deformed entangling surface. That is, we wish to introduce a `cylindrical' coordinate system that is centered on the deformed entangling surface. Such coordinates can be constructed perturbatively in the distance $\rho\equiv |x^a|$ from the entangling surface. We will use $K_{ij}^a$ to denote the extrinsic curvature of $f\Si$ and introduce the following notation for the trace and traceless parts:
\begin{equation}
K^a \equiv (K^a)_i{}^i\,,\qquad
\tilde{K}^a_{ij} \equiv  K^a_{ij} - \frac{ K^a}{d-2}\, \delta_{ij}\,.
\end{equation}
The new adapted coordinates are related to the previous Cartesian coordinates as follows:
\begin{equation}
\begin{array}{l}\label{change}
x'^a  =  x^a - f^a(y)- \frac{1}{d-2} \left( x^a K^b x_b - \frac{1}{2} K^a x^2 \right)
+O(\rho^4) \,, \\
y'^i  =  y^i + \pa^i f^a(y) x_a -\frac{1}{2(d-2)} x^2 \partial^i K^a x_a +O(\rho^5)\,,
\end{array}
\end{equation}
and the metric becomes (to reduce the clutter, we neglect the primes in the following but this metric is understood to be in the new adapted coordinate system)
\begin{align}\label{bdy_metric}
ds^2 & =
 \left(1+ \tfrac{2 K^c x_c}{d-2}  \right)\left( \rho^2 d\tau^2
+  d\rho^2 + [ \delta_{ij} +2  \tilde{K}^a_{ij}x_a ] dy^i dy^j + \tfrac{4}{d-2}   \partial_i  K^b\, x_b \rho  d\rho dy^i\right) + \mathcal{C}\,,
\end{align}
where ${x}^a=(\rho \cos\tau,\rho \sin\tau)$, and $\mathcal{C}$ represents the higher order terms with
\begin{equation}
\mathcal{C}=O(\rho^3)  d\rho^2  + O(\rho^5)  d\tau^2
+O(\rho^4) d\rho d\tau + O(\rho^4) \,  d\rho dy^i  +O(\rho^5)\,  d\tau dy^i
 +O(\rho^3)\, dy^i dy^j\, .
\end{equation}
Here, we consistently kept track of the corrections to the metric coming from any further change of coordinates allowed by symmetries and linear in the deformation $f^a$. We do not need to make any assumptions on those terms, because the order at which we work already allows to determine $C_D$.
Of course, the leading order in $\rho$ reduces to the well studied undeformed case. This is obvious from dimensional analysis, and will be useful in section \ref{hollow}.

Notice that the change of coordinates \eqref{change} simplifies for a traceless extrinsic curvature, \ie when $K^a=0$. When $d>3$, in order to determine $C_D$, it is sufficient to consider a deformation of this kind. (While it may not be possible to set $K^a=0$ everywhere, all of our calculations are local and so this does not matter). However, the choice of the frame defined by eq.~\eqref{change} has two advantages. It is convenient in $d=3$, where the extrinsic curvature has no traceless component.
Further, in higher dimensions, accommodating deformations for which $K^a$ is nonvanishing allows us to perform a consistency check on our computation of $C_D$, by considering both the traceless and trace contributions.

As a last step, we apply two consecutive Weyl transformations. The first with scale factor $\O_1 = (1- K^c x_c/(d-2))$ to remove the prefactor in the metric \eqref{bdy_metric} and the second with $\Omega_2={1}/{\rho}$ in anticipation of our holographic computations. After the first rescaling,\footnote{We emphasize that the rescaled metric no longer corresponds to flat space.} the metric exhibits an advantage of the change of coordinates \eqref{change}. Indeed if $f^a(y)$ implements a conformal transformation, eq.~\eqref{change} is the inverse transformation. In particular, starting from a planar defect, a conformal transformation maps it to a sphere, whose extrinsic curvature is simply $K^a_{ij}=\frac{1}{d-2} \delta_{ij} K^a$ with constant $K^a$ (\ie $\tilde K_{ij}^a=0$ and $\pa_i K^a=0$).  Hence, after the Weyl rescaling $f^a$ correctly appears in the metric only via $\tilde{K}_{ij}^a$ and derivatives of $K^a$, such that, for the map to a sphere, it would trivialize to the flat space metric. Furthermore the position of $\tilde K_{ij}^a$ is fixed by contraction of the indices, whereas the last term in the second line of eq.~\eqref{change} forces the trace of the extrinsic curvature to appear in as few places as possible.

The second Weyl rescaling with $\O_2=1/\rho$ does not provide an equivalent simplification, but it will turn out to be useful for the holographic computation in section \ref{hollow}. After the transformation $ G_{\mu\nu}\to  \frac{1}{\rho^2}\, G_{\mu\nu}$, we find
the conformally equivalent metric
\begin{equation}\label{metricdefhyp}
\begin{split}
ds^2 & =
  d\tau^2
+ \frac{1}{\rho^2} \left( d\rho^2 + [ \delta_{ij} +2 \, \tilde{K}^a_{ij}x_a ] dy^i dy^j + \frac{4}{d-2}   \partial_i \, K^b x_b \rho  d\rho dy^i \right) + \mathcal{C}' ,
\end{split}
\end{equation}
where the higher order corrections $\mathcal{C}'$ now take the form
\begin{equation}\label{metricdefhypC}
\mathcal{C}'= O(\rho)  d\rho^2  + O(\rho^3)  d\tau^2
+O(\rho^2) d\rho d\tau + O(\rho^2) \,  d\rho dy^i  +O(\rho^3)\,  d\tau dy^i
 +O(\rho)\, dy^i dy^j .
\end{equation}
The metric above
describes a slightly deformed version of the manifold $S^1 \times H^{d-1}$, appearing \eg in \cite{Casini:2011kv,Hung:2011nu} -- see also section \ref{hollow}. In particular, the deformation decays asymptotically as we approach the asymptotic boundary of the hyperbolic hyperplane with $\rho\to0$. We denote the new geometry as $\tilde H_n$.

In these coordinates, the stress tensor one-point function looks particularly simple. In order to write it down in even dimensions, we should be careful to include the effect of the conformal anomaly. Under the rescaling $ G_{\mu\nu}\to \widetilde G_{\mu\nu} = \O^2 G_{\mu\nu}$ (here $\Omega=\Omega_1 \Omega_2$), the stress tensor one-point function transforms as follows:
\begin{equation}\label{Weylan}
 \braket{\widetilde {T}_{\mu\nu}}_n
 =\O^{2-d} \braket{T_{\mu\nu}}_n
 +\mathcal{A}_{\mu\nu},
\end{equation}
where $\langle \widetilde T_{\mu\nu} \rangle_n$ is the stress tensor expectation value after the rescaling. The anomalous contributions $\mathcal{A}_{\mu\nu}$ are the higher dimensional analog of the Schwarzian derivative appearing in $d=2$ and are independent of $n$, because locally the $n$--fold branched cover is identical to the original spacetime manifold \cite{Hung:2011nu,Hung:2014npa}.
It is therefore possible to subtract this contribution without knowing its explicit form. Using the fact that in flat space the vacuum expectation value of the stress tensor vanishes, \ie $\braket{T_{\mu\nu}}_{n=1}=0$, one easily finds
\begin{equation}
 \mathcal{A}_{\mu\nu}=\braket{\widetilde {T}_{\mu\nu}}_1 .
\end{equation}
Therefore combining the above results, we can write
\begin{equation}\label{Tgoodframe}
\begin{split}
\langle \widetilde T_{ab} (x) \rangle_n
& = \frac{g_n}{\rho^2} \left((d-1) \delta_{ab} - d \frac{ x_a x_b}{\rho^2}\right) +\ldots
,
\\
\langle \widetilde T_{ai} (x) \rangle_n
& = \frac{x_a x_b }{\rho^2}\del_i K^b \frac{k_n}{d-2}  + \ldots   ,
\\
\langle \widetilde T_{ij} (x) \rangle_n
& = \frac{1}{\rho^2} \left( -g_n \delta_{ij} + k_n \tilde K^a_{ij} x_a \right) +\ldots,
\end{split}
\end{equation}
where
\begin{align}\label{eq:CFToutputkn}
k_n-k_1&=\frac{(d-1)   \Gamma \left(\frac{d}{2}-1\right)\pi ^{\frac{d}{2}-2}}{2\Gamma (d+1)}\frac{C_D}{n}-\frac{3d-4}{d-2} \frac{h_n}{2\pi n}\, &
g_n-g_1&=\frac{h_n}{2\pi n}\,.
\end{align}
\rev{\emph{En passant}, we note that $k_n-k_1 = -(g_n-g_1)$ when the conjecture is satisfied.}
We emphasize that the anomalous contributions only appear in even dimensions and hence $k_1$ and $g_1$ vanish in odd dimensions. The ellipses stand for higher orders in $\rho$, which are the same as in the metric \eqref{metricdefhyp} when written in component form.
Eq.~\eqref{Tgoodframe}, together with the metric \eqref{bdy_metric}, are the only ingredients entering the holographic computation.
Let us also point out that the one-point function \eqref{Tgoodframe} and the metric \eqref{metricdefhyp} have similar structure in terms of the extrinsic curvature. In view of holographic renormalization, this suggests that the bulk metric will preserve the simplicity of the boundary metric.

\section{Shape Deformations from Holography}\label{hollow}

In this section, we use holography to compute the one-point function of the stress tensor and then compare the holographic results  to the field theoretic expressions in eqs.~\eqref{Tgoodframe}-\eqref{eq:CFToutputkn} in order to extract $C_{D}$.

As mentioned at the beginning of section \ref{cftside}, the \ren\ entropy can be evaluated using the partition function $Z_n$ on a branched $n$-fold cover of the original $d$-dimensional spacetime. Implicitly, the latter path integral can be used to define the twist operator using eq.~\reef{sloppy}. For the purposes of our holographic calculations, it turns out that it is most convenient to work with this geometric interpretation. In particular, we will be extending the holographic computations introduced in \cite{Casini:2011kv,Hung:2011nu}. The discussion there began by considering how to evaluate the entanglement and \ren\ entropies for a spherical or flat entangling surface in the flat space vacuum of a general CFT. By employing an appropriate conformal transformation, this question was then related to understanding the thermal behaviour of the CFT on a hyperbolic hyperplane. That is, the partition function $Z_n$ was conformally mapped to the Euclidean path integral on the geometry $S^1\times H^{d-1}$, \ie the product of a periodic Euclidean circle and a $(d-1)$-dimensional hyperbolic space. Next in the case of a holographic CFT, this thermal partition function is evaluated by considering a so-called `topological' AdS black hole with a hyperbolic horizons. In fact, the latter solutions can be found for a variety of higher derivative theories, as well as Einstein gravity \cite{Hung:2011nu,brandon}.

An important element of the conformal mapping in \cite{Hung:2011nu} is that
the conical singularity at the entangling surface in the branched cover of flat  space is `unwound' by extending the periodicity on the thermal circle. To make this statement precise, let us consider the metric in eq.~\eqref{metricdefhyp} for the undeformed case, \ie with $K_{ij}^a=0$. In this case, the geometry is precisely
$S^1\times H^{d-1}$ with the radius of curvature on $H^{d-1}$ implicitly set to one. Further, beginning with a $n$-fold cover of flat space, the periodicity of the $\tau$ circle is $\tau\sim \tau+2\pi n$. Considering the path integral of the CFT on this background then yields the corresponding thermal partition function with temperature $T = 1/ (2 \pi n)$. However, the important point is that this boundary geometry is completely smooth, which makes the question of finding the dual bulk configuration relatively straightforward.  Of course, as noted above, the desired bulk solution corresponds to a hyperbolic black hole in AdS space.

The problem which we face then is to extend this holographic analysis to accommodate  deformations away from the very symmetric entangling surfaces considered in the calculations described above. For a generic deformation of a flat or spherical entangling surface, the dual bulk geometry is not known, but the question of small deformations is precisely the one addressed by \cite{Dong:2016wcf} in $d=4$. Hence we must only extend this analysis to general dimensions. In fact, we also extend these calculations to a broader class of shape deformations. An essential feature of this approach is that we only solve for the bulk geometry at leading order in the size of the deformation of the entangling surface in the boundary.

With a small deformation, we can solve for the bulk geometry order by order in the distance from the entangling surface $\rho$. The leading order solution coincides with the black hole geometry described above for an undeformed entangling surface. One can then move to the next order in $\rho$ to compute the bulk metric at first order in the deformation. Once our bulk metric is determined, we can extract $C_D$. As explained in the introduction, our procedure involves computing the one-point function of the stress tensor, to enhance the appearance of $C_D$ to leading order in the deformation. This one-point function will  be computed using standard holographic renormalization techniques \cite{deHaro:2000vlm}.

\subsection{Holographic Setup}
\label{holo}

Let us start by introducing an ansatz for the bulk metric. We first observe that the parallel components of the metric \eqref{metricdefhyp} only depend on the traceless part of the extrinsic curvature, while the $g_{\rho i}$ components contain contributions from the parallel derivatives of the trace of the extrinsic curvature. This was achieved by our choice of coordinates \eqref{change} (plus the Weyl rescaling) and it is convenient in minimizing the number of unknown functions required for the gravitational ansatz. The bulk metric can then be written as
\begin{equation}
\begin{split}
ds_{bulk}^2 = & \frac{dr^2}{\frac{r^2}{L^2}g(r)-1} + \left( \frac{r^2}{L^2} g(r) -1   \right) L^2 d\tau^2
\\
&+ \frac{r^2}{\rho^2} \left( d\rho^2 + [ \delta_{ij} +2 \, k(r) \tilde{K}^a_{ij}x_a ] dy^i dy^j + \frac{4}{d-2} v(r)  \partial_i \, K^b x_b \rho  d\rho dy^i \right) + \cdots ,
\end{split}
\label{metric}
\end{equation}
where again the ellipsis stand for higher orders in $\rho$, and $L$ denotes the AdS curvature scale.
We will refer to the functions $k(r)$ and $v(r)$ as the
traceless and traceful parts of the gravity solution, respectively. Their value will be determined by solving gravitational equations of motion at the first subleading order in $\rho$. This procedure produces two second-order differential equations for $k(r)$ and $v(r)$ which we must solve numerically for general values of $n$. We are also able to obtain analytic solutions in the vicinity of $n=1$, as well as $n\to0$.
As boundary conditions, we require $k(r)\rightarrow 1$ and $v(r) \rightarrow 1$ as we approach the AdS boundary ($r\to\infty$) to reproduce the desired boundary metric. We also demand that the geometry is smooth at the `horizon', \ie where $g_{\tau\tau}$ vanishes.

\subsection{Einstein Gravity}\label{Einstein}
In this subsection, we extract $C_D$ for the boundary theories whose holographic dual is described by Einstein gravity.
The metric function $g(r)$, which is determined by the Einstein equations at zeroth order in $\rho$, is given by \cite{Hung:2011nu,Dong:2016wcf}
\begin{align}\label{eq:black_factor}
g(r) &= 1 -\frac{r_h^d - L^2 r_h^{d-2}}{r^{d}},
\end{align}
where $r_h$ is the position of the horizon (in Lorentzian signature). It will be useful to define the dimensionless variable $x_n \equiv r_h/L$. Then $x_n$ is related to $n$ by
\begin{eqnarray}
n = \frac{2 x_n}{d \left(x_n^2-1\right)+2} \,.\label{obtuse}
\end{eqnarray}

At the next order in $\rho$, the Einstein equations yield a second order differential equation for $k(r)$,
\begin{equation}
k''(r) + \frac{r^3 g'(r) + (d+1) r^2 g(r)-(d-1) L^2}{r^3 g(r)-L^2 r} k'(r) - \frac{L^2 \left((d-3) \left(r^2 g(r)-L^2\right)+r^2\right)}{\left(L^2 r-r^3 g(r)\right)^2} k(r) =0 \,, \label{eqEOMr}
\end{equation}
as well as the algebraic equation
\begin{equation}\label{nu_equal_k}
k(r) = v (r) \,.
\end{equation}
Note that for $d=4$, eq.~\reef{eqEOMr} correctly reproduces the analogous equation appearing in \cite{Dong:2016wcf}.
Other components of the Einstein equations give additional first and second order equations for $v(r)$, which are automatically solved when eqs.~\eqref{eqEOMr} and \eqref{nu_equal_k} are satisfied. To derive eq.~\eqref{nu_equal_k}, we used the Gauss-Codazzi relations $\del_k K_{ij}^a = \del_j K_{ik}^a$ (at leading order in $f^a$).
The equality \eqref{nu_equal_k} provides a nontrivial consistency check of our ansatz \eqref{metric} for the bulk metric. Indeed, eq.~\eqref{Tgoodframe} shows that $C_D$ and $h_n$ appear in the same combination, denoted $k_n$, in factors multiplying both the traceless and the traceful parts of the deformation. Eq.~\eqref{nu_equal_k} ensures that the holographic solution will match this prediction from the CFT.
The case of $d=3$ is slightly different since the traceless part of the extrinsic curvature $\tilde{K}^a_{ij}$ vanishes. We therefore find that Einstein equations contain only the second order differential equation for $v(r)$
\begin{eqnarray}
v''(r) + \frac{r^3 g'(r)+4 r^2 g(r)-2 L^2}{r^3 g(r)-L^2 r} v'(r) - \frac{L^2 r^2}{\left(L^2 r-r^3 g(r)\right)^2} v(r) = 0 \,, \label{eqEOMvd3}
\end{eqnarray}
which matches eq.~\eqref{eqEOMr} upon substituting $v(r)=k(r)$ and $d=3$.

\subsubsection{Holographic Renormalization}\label{holgra}
Given the bulk metric \eqref{metric}, we are interested in evaluating the boundary expectation value of the stress tensor. This computation can be performed using the technique described in \cite{deHaro:2000vlm}. First, we write the metric in the Fefferman-Graham (FG) form~\cite{Fefferman:2007rka}
\begin{equation}
ds^2_{bulk} = \frac{L^2}{z^2}\left(dz^2 + h_{\mu \nu}(x,z)\, dx^\mu dx^\nu\right) \,, \label{FGmetric}
\end{equation}
where
\begin{equation}
h_{\mu\nu}(x,z) = h_{(0) \mu\nu}(x) + z^2\, h_{(2)\mu\nu}(x) +
\cdots + z^d\, h_{(d)\mu\nu}(x) + \cdots \,.
\label{smallz}
\end{equation}
The expectation value for the stress tensor is then determined by the $h_{(i)}$'s, with the following general expression
\begin{equation}\label{holT}
\vev{T_{\mu\nu}}_{\tilde{H}_n} = \frac{d}{2}\left(\frac{L} {\lp}\right)^{d-1} h_{(d)\mu\nu}+ \mathcal{X}_{\mu\nu}\left[h_{(m)\mu\nu }\right]_{m<d}.
\end{equation}
The subscript $\tilde{H}_n$ indicates that the expectation value is taken in the deformed boundary geometry described by eq.~\reef{metricdefhyp}. Here $\mathcal{X}_{\mu\nu}$ is a functional of the lower order $h_{(i)}$ terms, which are completely fixed by the boundary geometry. This contribution is related to the Weyl anomaly and accordingly, it vanishes with an odd number of boundary dimensions. In even $d$, its explicit expression depends on the dimension. For the cases  $d=4$ and $6$, the interested reader is referred to eqs.~(3.15) and (3.16) in \cite{deHaro:2000vlm}. We will see that it is not necessary to compute those contributions in order to obtain $C_D$. However, for completeness, we show how to obtain the exact expressions for the expectation value of the stress tensor in appendix \ref{app_holo_ren}.

By comparing eqs.~\eqref{holT} and \eqref{metric} with eq.~\eqref{Tgoodframe}, we see that the expansions of $k(r)$ and $v(r)$ near the boundary carry the information about the displacement operator. In this limit, the form of the solution to the equations of motion \eqref{eqEOMr}-\eqref{nu_equal_k} reads
\begin{equation}
\begin{split}\label{eq:k_r_expansion}
d  = & \, 3\quad , \quad k(r)=v(r)=1-\frac{L^2}{2r^2} + \frac{L^3}{r^3} \beta_n +\ldots ,  \\
d = & \, 4 \quad , \quad k(r)=v(r)=1-\frac{L^2}{2r^2} + \frac{L^4 }{r^4} \beta_n +\ldots, \\
d  = & 5 \quad , \quad k(r)=v(r)=1-\frac{L^2}{2r^2} - \frac{L^4}{8r^4} + \frac{L^5}{r^5} \beta_n +\ldots, \\
d  = & 6 \quad , \quad k(r)=v(r)=1-\frac{L^2}{2r^2} - \frac{L^4}{8r^4} + \frac{L^6}{r^6} \beta_n +\ldots .
\end{split}
\end{equation}
Here, $\beta_n$ is the first coefficient which is not fixed by the boundary conditions at infinity. As one might expect, this coefficient determines $C_D$, and we obtain it numerically in the next subsection.
Matching these expansions with eq.~\eqref{Tgoodframe}, we find the following relations:
\begin{equation}
\begin{split}\label{kghol}
k_n&=\left(\frac{L}{\lp} \right)^{d-1} \left( x_n^d-x_n^{d-2}+d\beta_n + k_0^{(d)} \right)  \,,
\\
g_n&=- \left(\frac{L}{\lp} \right)^{d-1} \left( \frac{x_n^d-x_n^{d-2}+g_0^{(d)}}{2} \right) \,,
\end{split}
\end{equation}
where $k_0$ and $g_0$ contain the anomalous contributions. As mentioned before, these vanish for odd dimensions and are independent of $n$ in even dimensions\footnote{In particular, one can find that for $d=4$, $k_0^{d=4} = 3/4$ and $g_0^{d=4}=1/4$, and for $d=6$, $k_0^{d=6} = 5/8$ and $g_0^{d=6}=-3/8$.\label{footnote}} -- see appendix \ref{app_holo_ren}. Note that in order to obtain $C_D$ and $h_n$ from eq.~\eqref{eq:CFToutputkn}, we only need to consider the differences $k_n-k_1$ and $g_n-g_1$. Then, all the anomalous contributions will cancel.\footnote{Notice that in our conventions the stress tensor has lowered indexes, contrary to the one in
\cite{Dong:2016wcf}. The dictionary between the two conventions is as follows: $P_n = -g_n$ and $\alpha_n = k_n+ 4 g_n$, with $\lp^{d-1}=8\pi G_N$. This gives precise agreement between both expressions in $d=4$.}

Comparing eqs.~\eqref{kghol} and \eqref{eq:CFToutputkn}, we find holographic expressions for $C_D$ and $h_n$,
\begin{eqnarray}
\frac{h_n}{\pi n} & = & \left(\frac{L}{\lp} \right)^{d-1} \left( x_n^{d-2}-x_n^d \right) \,, \\
\frac{C_D}{n} & = &  \frac{d\,\Gamma(d+1) }{(d-1)\pi^{d/2-2}\Gamma({d}/{2})}
\left((d-2)\left(\frac{L}{\lp} \right)^{d-1} (\beta_n-\beta_1)
+\frac{h_n}{2\pi n}\right) \,.\label{truse}
\end{eqnarray}
The Planck length $\lp$ can be replaced for CFT data as follows, \eg see \cite{Hung:2014npa}:
\begin{eqnarray}
C_T = \left(\frac{L}{\lp} \right)^{d-1} \left( 2^{d-2}  \pi ^{-\frac{d+1}{2}}d (d+1) \Gamma \left(\frac{d-1}{2}\right) \right) \,,
\end{eqnarray}
where $C_T$ is the coefficient that appears in the two-point function of the vacuum stress tensor \cite{Osborn:1993cr,Erdmenger:1996yc},
\beq
\vev{T_{\mu\nu}(x) T_{\rho\sigma}(0)} = \frac{C_T}{x^{2d}}{\cal{I}}_{\mu\nu,\rho\sigma}(x)\,.
\label{ozzy}
\eeq

In order to obtain $C_D$, we now only need to solve numerically the equations of motion \eqref{eqEOMr} and extract $\beta_n$.
We will compare $C_D$ with the value in eq.~\reef{conj} related to previous conjectures \cite{Bianchi:2015liz}
\begin{equation}
C_D^{\conj}(n) = d \, \Gamma\left(\frac{d+1}{2}\right)\left( \frac{2}{\sqrt{\pi}}\right)^{d-1}h_n \,. \tag{\ref{conj}}
\end{equation}
We will find that the conjecture is violated for holographic theories in any spacetime dimension. This conclusion will be supported numerically for $3\leq d\leq6$ with arbitrary $n$ in section \ref{num}, and also with analytic results near $n=0,1$ in general dimensions in section \ref{ana}. In particular, the expected agreement with eq.~\eqref{conj} is reproduced only at linear order in $(n-1)$, but we see $C_D$ will depart from eq.~\reef{conj} at order $(n-1)^2$.

\subsubsection{Numerical Solutions}\label{num}
To solve the second order differential equation \eqref{eqEOMr}, we use a shooting method. The two integration constants will be free coefficients in the asymptotic expansions near both limits of integration. Near the asymptotic boundary, we have $\beta_n$ while regularity of the solution near the horizon fixes a new integration constant. In particular, near the horizon we need $k(r) \propto (r/L-x_n)^{n/2}$, where the proportionality constant will provide the second integration constant. It is useful to consider coordinates in which the extreme values are kept fixed. Hence for our numerical integrations, we defined $\tilde{r} \equiv (x_n L)/r$, so that the AdS boundary is at $\tilde{r}_{bdy}=0$ and the horizon, at $\tilde{r}_{hor} = 1$. For each value of $n$, we solve the equation numerically both from the boundary and the horizon, fixing the integration constants so that the two curves meet smoothly.

\begin{figure}[htb]
\setlength{\abovecaptionskip}{0 pt}
\centering
\includegraphics[scale=0.6]{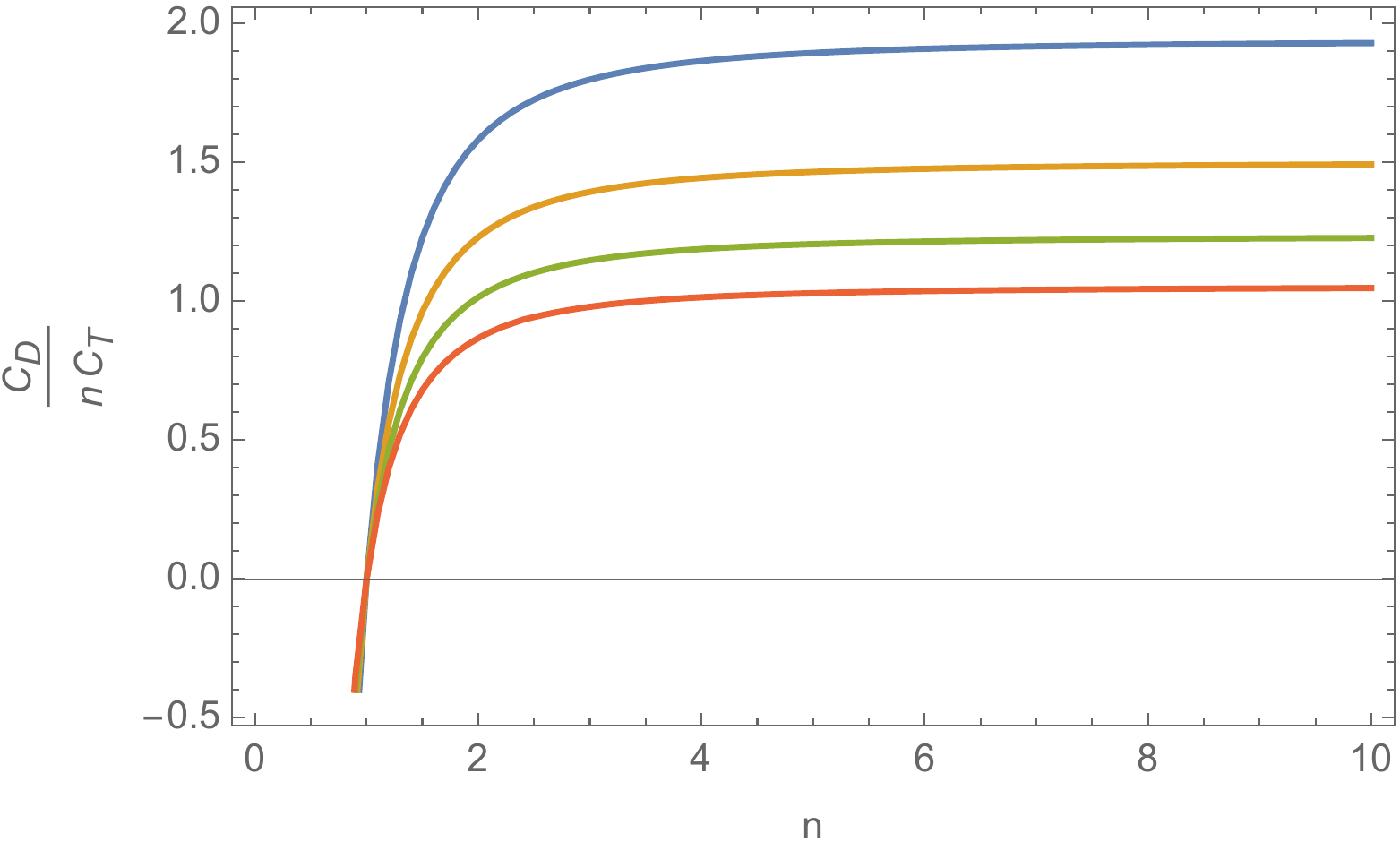}
\caption{$C_D/(n C_T)$ as a function of $n$. Different curves correspond to $d=3$ (blue), $d=4$ (yellow), $d=5$ (green) and $d=6$ (red).} \label{fig_cd}
\end{figure}

\begin{figure}[htb]
\setlength{\abovecaptionskip}{0 pt}
\centering
\includegraphics[scale=0.75]{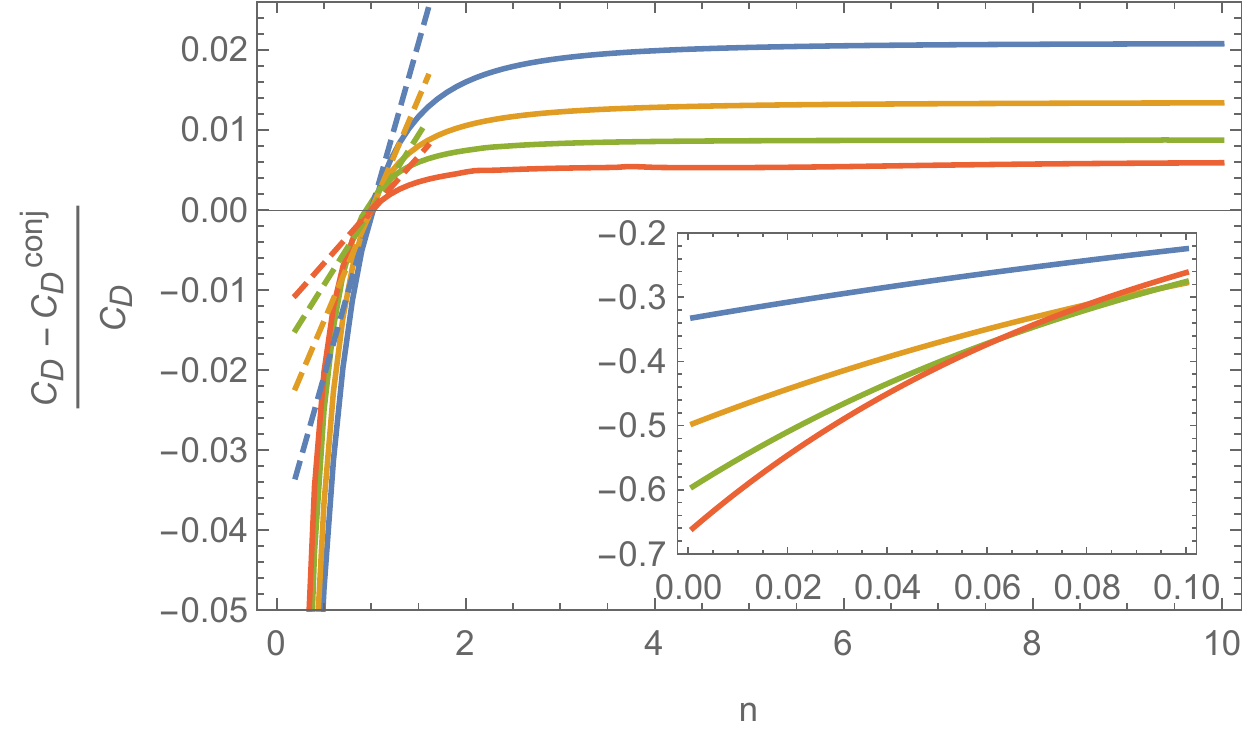}
\caption{Relative mismatch between $C_D$ and the conjectural value \eqref{conj} as a function of $n$ for $d=3$ (blue), $d=4$ (yellow), $d=5$ (green) and $d=6$ (red). Dashed lines show the leading order analytic solution around $n=1$, supporting the numerical data. In the inset, we show the numerical results near $n=0$, which smoothly approach the value $(2-d)/d$ at $n=0$, as predicted analytically in eq.~\eqref{rel_near0}.} \label{fig_rel}
\end{figure}

The results for $C_D$ are plotted in fig.~\ref{fig_cd}. In the figure, we chose to normalize $C_D$ by a factor $n$, in order to exhibit that this combination reaches a fixed value at large values of the \ren\ index. Notice that, due to the prefactor in the definition of the \ren\ entropies \eqref{renyidef}, this normalization quantifies more precisely the shape dependence of $S_n$ at large $n$.
As one can see from fig.~\ref{fig_rel}, $C_D$ deviates from $C_D^{\conj}$ away from the linear regime around $n=1$. Yet, notice that curiously, the relative difference $\frac{C_D-C_D^{\conj}}{C_D}$ is fairly small for all $n>1$. Although we are sure that this difference is bigger than our numerical accuracy, the analytic solution of the differential equation \eqref{eqEOMr} close to $n=1$ confirms that eq.~\reef{conj} fails (for general dimensions), as does the analytic result for the limit $n\to0$.

\subsubsection{Analytic Solutions}\label{ana}
It is also possible to produce an analytic treatment of eq.~\eqref{eqEOMr} near $n=1$.
We can solve the equation analytically order by order in powers of $(n-1)$ and then fix the integration constants by providing the boundary expansion for $k(r)$ and regularity near the horizon. We find that
\begin{equation}
k(\tilde{r}) = k_0(\tilde{r}) + k_1(\tilde{r}) (n-1) + k_2(\tilde{r}) (n-1)^2 + O(n-1)^3 \,,
\end{equation}
with
\begin{equation}
\begin{split}
k_0(\tilde{r})&=\sqrt{1-\tilde{r}^2} \,,
\\
k_1(\tilde{r})&=\frac{(d-1) \left(\tilde{r}^2-1\right) \tilde{r}^d \,\, _2F_1\left(1,\frac{d}{2};\frac{d+2}{2};\tilde{r}^2\right)+d \left(\tilde{r}^d-\tilde{r}^2\right)}{(d-1) d \sqrt{1-\tilde{r}^2}} \,.
\end{split}
\end{equation}
For $k_2(x)$ we solve separately for each dimension. These results determine  $\beta_n$ perturbatively around $n=1$, and the result can be written as
\begin{align}\label{betaan}
\beta_n^d=
\beta_1^d
+\frac{1}{d(d-1)}(n-1)-\frac{4 d^3 -8 d^2+d+2}{2\,d^2 (d-1)^3 }(n-1)^2 + O(n-1)^3 \,,
\end{align}
with $\beta_1^d$ being zero for odd $d$ and $\beta_1^d = -\frac{\Gamma \left(\frac{d-1}{2}\right)}{2 \sqrt{\pi } \Gamma \left(\frac{d}{2}+1\right)}$ for even $d$.

Given this expansion for $\beta_n$ and the corresponding expansion for $x_n$ from eq.~\reef{obtuse}, it is straightforward to compute $C_D$ as a power series in $(n-1)$:
\begin{equation}
\frac{C_D}{C_T} = \frac{2 \pi ^2 }{d+1}(n-1)-\frac{2 \pi ^2 \left(d^2-d-1\right) }{d^3-d}(n-1)^2 + O(n-1)^3 \,,
\end{equation}
which, as expected \cite{Faulkner:2015csl}, agrees with the conjecture \reef{conj} at linear order,
\begin{equation}
\frac{C_D^{\conj}}{C_T}=\frac{2 \pi ^2 }{d+1}(n-1)-\frac{\pi ^2 (2 d^2-4d +1) }{(d-1)^2 (d+1)}(n-1)^2 + O(n-1)^3 \,,
\end{equation}
but not at second order. In fact, the relative mismatch between the two expressions can be easily computed and is given by
\begin{equation}
\frac{C_D-C_D^{\conj}}{C_D}=\frac{(d-2) }{2\,d (d-1)^2}(n-1)+O\left(n-1\right)^2.
\end{equation}

\
\

Interestingly, one can also extract the analytic expression for $C_D$ at leading order as $n\to0$. This result follows from the observation that the $\beta_n$ contribution in eq.~\reef{truse} is subleading with respect to $x_n^d$ at small $n$. More precisely, one can verify that $\beta_n/x_n^d \sim n$ in this limit. Then, we do not actually need to solve eq.~\eqref{eqEOMr} but just expand $x_n$ for small $n$ to find
\begin{eqnarray}
\frac{C_D}{C_T} & = &
-\left(\frac{1}{dn}\right)^{d-1}  \left( \frac{2^{d-1} \pi ^2 }{d+1} + O(n)\right)
\,, \\
\frac{C_D^{\conj}}{C_T} & = &
-\left(\frac{1}{d n}\right)^{d-1} \left(\frac{2^d\pi ^2  (d-1)}{d(d+1)} + O(n)\right)\,,
\end{eqnarray}
which yields
\begin{equation}
\frac{C_D-C_D^{\conj}}{C_D}  = - \frac{d-2}{d} + O(n) \,. \label{rel_near0}
\end{equation}
Note that the relative error is order one as $n$ goes to zero, contrary to the small differences which were obtained for $n>1$.

\subsection{Gauss-Bonnet Gravity}\label{gaussbonnet}
In this section, we consider holographic CFTs dual to Gauss-Bonnet (GB) gravity. The full gravitational action reads \cite{Buchel:2009sk}
\begin{equation}
 I=\frac{1}{2 \ell_p^{d-1}} \int d^{d+1} x \sqrt{-g} \left[\frac{d(d-1)}{L^2}+R+\frac{\lambda\, L^2}{(d-2)(d-3)} \mathcal{X}_4 \right], \label{gamma}
\end{equation}
where
\begin{equation}\label{GBaddterm}
 \mathcal{X}_4=R_{abcd}R^{abcd}-4R_{ab} R^{ab}+R^2,
\end{equation}
and the term \eqref{GBaddterm} contributes to the equations of motion only for $d\geq 4$ (note that the bulk theory is  $d+1$ dimensional).
The coupling $\lambda$ is constrained by known unitarity bounds \cite{Buchel:2009sk}
\begin{equation}\label{bounds}
-\frac{(3d+2)(d-2)}{4(d+2)^2}\leq\lambda \leq \frac{(d-2)(d-3)(d^2-d+6)}{4(d^2-3d+6)^2} \,.
\end{equation}
The same constraints can also be derived by excluding the propagation of superluminal modes in thermal backgrounds \cite{wok1,wok2,wok3}. Before proceeding, we should add a word of caution since in fact a detailed analysis indicates that the GB theory \reef{gamma} violates causality unless the spectrum is supplemented by some higher spin modes \cite{Camanho:2014apa}. However, it remains unclear in which situations these additional degrees of freedom will play an important role. Hence we proceed with the perspective that these holographic theories are amenable to simple calculations and allow us to investigate a broader class of holographic theories. Further, such investigations may still yield interesting insights on universal properties which may hold for general CFTs, beyond the holographic CFTs defined by these toy models. Certainly, this approach has been successful in the past,  \eg in the discovery of the F-theorem \cite{Myers:2010xs,Myers:2010tj}.

Conceptually, the procedure here is completely analogous 
to the one for the Einstein gravity case analyzed in the previous section, although explicit computations can become more tedious due to the $\lambda$-dependence.

In order to have the appropriate AdS asymptotics, we slightly modify the bulk metric ansatz,
\bea
ds_{bulk}^2 = \frac{dr^2}{\frac{r^2}{L^2}g(r)-1}& + &\left(\frac{r^2}{L^2}g(r)-1\right) \frac{L^2}{g_{\infty}}\, d\tau^2 \label{house}
\\
&&\qquad+ \frac{r^2}{\rho^2} \left( d\rho^2 + [ \delta_{ij} +2 \, k(r) \tilde{K}_{aij}x^a ] dy^i dy^j \right) + \cdots \,.  \nonumber
\eea
Note the additional factor of $g_\infty$ in the $\tau\tau$ component, which is defined below. The metric for the hyperbolic black holes in GB gravity reads, \eg \cite{brandon}
\begin{equation}
g(r) =\frac{1}{2\lambda} \left[1-\sqrt{1-4\lambda
\left(1- \frac{r_h^d -L^2 r_h^{d-2} +\lambda L^4 r_h^{d-4}}{r^d}\right)}\right] \,.\label{end}
\end{equation}
It is useful to define the asymptotic limit of $g(r)$ as $r$ goes to infinity,
\begin{equation}
g_{\infty} \equiv \lim_{r\to\infty} g(r) =  \frac{1-\sqrt{1-4\lambda}}{2\lambda} \,.\label{end2}
\end{equation}
Now we observe that the AdS curvature scale is no longer simply given by $L$, the scale appearing in the action \reef{gamma}.  Instead the AdS scale becomes $\tilde L = L / \sqrt{g_{\infty}}$, as can be seen by examining the asymptotic limit of  $g_{rr}$ in eq.~\reef{house}. Hence we find it more (physically) convenient to write our expressions for GB gravity in terms of $\tilde L$, rather than $L$. The next step is to relate the position $r_h$ of the black hole horizon to the R\'enyi index $n$. In the GB gravity case, this relation is more complicated than with Einstein gravity. In particular, it is given implicitly by \cite{Hung:2011nu}
\begin{equation}
0=(d-4) g_{\infty} \lambda +\frac{ (4 g_{\infty} \lambda )}{n}x_n -(d-2) x_n^2-\frac{2 }{n}x_n^3+\frac{d }{g_{\infty}}x_n^4 \,,
\end{equation}
where now we have redefined $x_n \equiv r_h/\tilde L$. 
For simplicity, we restrict the following analysis to considering traceless deformations, \ie with $K^a=0$, as this is enough to extract $C_D$ in any $d\geq 4$. Note that the results for the GB theory will only differ from those for Einstein gravity in that range of dimensions.\footnote{With $d=3$, \ie four dimensions in the bulk, the GB interaction \reef{GBaddterm} becomes topological and does not modify the gravitational equations of motion.}

As before, the expression for $g(r)$ given in eq.~\reef{end} solves the gravitational equations at the leading order in $\rho$. The first subleading order in $\rho$ then provides the following equation for $k(r)$:
\begin{equation}
a(r)k''(r)+b(r)k'(r)+c(r)k(r)=0,
\end{equation}
where
\begin{align}
\begin{split}
a(r)=& \ r^2 \left(\tilde{L}^2 g_{\infty} -r^2 g(r)\right) \left(\frac{2 \lambda }{d-2} r g'(r)+2 \lambda  g(r)-1\right),
\end{split}
\\
\begin{split}
 b(r)=& \ \frac{1}{d-2}
 \left(
2 \lambda  r^3 \left(\tilde{L}^2 g_{\infty}-r^2 g(r)\right) g''(r)-2 \lambda  r^5 g'(r)^2\right.
\\
&
+r^2 g'(r) \left(2 (2-3 d) \lambda  r^2 g(r)+4 (d-1) \lambda  \tilde{L}^2 g_{\infty}+(d-2) r^2\right)
\\
&\left.+(d-2) r (2 \lambda  g(r)-1) \left((d-1) \tilde{L}^2 g_{\infty}-(d+1) r^2 g(r)\right)\right),
\end{split}
\\
\begin{split}
c(r)=&\
g''(r)\frac{ 2 (d-2) \lambda  r^4 g(r)-(d-2) r^4+2 \lambda  \tilde{L}^2 g_{\infty} r^2}{d-2}+2 \lambda  r^4 g'(r)^2
\\
&+g'(r)\left(2 \lambda  \tilde{L}^2 g_{\infty} r \left(\frac{g_{\infty} r^2}{(d-2) \left(r^2 g(r)-\tilde{L}^2 g_{\infty}\right)}+2\right)+4 d \lambda  r^3 g(r)-2 d r^3\right)
\\
&+g(r) \left(2 (d-3) \lambda  \tilde{L}^2 g_{\infty}-(d-1) d r^2\right)+(d-1) d \lambda  r^2 g(r)^2-d \tilde{L}^2 g_{\infty}-d r^2
\\&
+\frac{ \tilde{L}^2 g^2_{\infty} \left(r^2-2 \lambda  \tilde{L}^2 g_{\infty}\right)}{\tilde{L}^2 g_{\infty}-r^2 g(r)}+2 g_{\infty}^2 \lambda  \tilde{L}^2+3 \tilde{L}^2 g_{\infty}+d^2 r^2\, .
\end{split}
\end{align}

As in eq.~\reef{eq:k_r_expansion} for Einstein gravity, the solution has a near-boundary expansion of the form
\begin{equation}
\begin{split}
d = & \, 4 \quad , \quad k(r)=1-\frac{\tilde L^2}{2r^2} + \frac{\tilde L^4}{r^4}\beta_n +\ldots, \\
d  = & 5 \quad , \quad k(r)=1 - \frac{\tilde L^2}{2r^2} - \frac{\tilde L^4}{8r^4} + \frac{\tilde L^5}{r^5}\beta_n +\ldots, \\
d  = & 6 \quad , \quad k(r)=1-\frac{\tilde L^2}{2r^2} - \frac{\tilde L^4}{8r^4} + \frac{\tilde L^6}{r^6}\beta_n+\ldots \,,
\end{split}
\end{equation}
and our task is to determine $\be_n$ in order to extract $C_D$. First, we evaluate $\be_n$  numerically in $d=4$ and 5 for arbitrary $n$. Then we also determine $\be_n$ analytically in an expansion about $n=1$ and at $n=0$, in $d=4,5$ and 6.

The one-point function of the stress-tensor is obtained as before via holographic renormalization (see section \ref{holgra}) \cite{Faulkner:2013ica}:\footnote{Note that eq.~(6.29) in \cite{Faulkner:2013ica}, which gives the expectation value for the stress tensor in arbitrary $R^2$ gravity, has a missing factor of 2 in the $a_1$-term which we have corrected here to obtain eq.~\reef{gorp}. Also note that the conventions for $L$ and $\tilde L$ are interchanged there. We follow the conventions in \cite{Hung:2011nu}.}
\begin{equation}
\langle T_{\mu\nu} \rangle = \frac{d\tilde L^{d-1}}{2 \lp^{d-1}} \left[1-2 \lambda g_\infty \right] h_{\mu\nu}^{(d)}+ \mathcal{X}_{\mu\nu}\left[h_{(m)\mu\nu }\right]_{m<d}\,.\label{gorp}
\end{equation}
where $\mathcal{X}_{\mu\nu}$ is again some functional of the lower order terms in the metric expansion and is related to the Weyl anomaly. As in the Einstein case, we will not need to compute those contributions in order to obtain $h_n$ or $C_D$. Note that in this case the FG expansion is exactly as in the Einstein case with the obvious difference that $L$ is replaced by $\tilde L$, \ie $ds^2_{bulk} =  \, \frac{\tilde L^2}{z^2}\left(dz^2 + h_{\mu \nu} dx^\mu dx^\nu\right)$.

Now we can evaluate $h_n$ in the boundary CFT dual to GB gravity case by examining the $g_n$ term in $\vev{T_{ij}}$ in eq.~\reef{Tgoodframe}. We recover the known result \cite{Hung:2011nu},
\begin{equation}\label{hnGB}
\begin{split}
h_n = \frac{1}{4} \Gamma \left(\frac{d}{2}\right) &\pi^{1-\frac{d}{2}} n \, x_n^{d-4} (x_n^2-1)
\\ &\times\ \left( (d-3) (x_n^2-1) a_d^*+(d-3-(d+1)x_n^2) \frac{(d-1)}{(d+1)}\frac{\pi^d}{\Gamma(d+1)} C_T \right),
\end{split}
\end{equation}
that we expressed in terms of\footnote{Note that we use here $C_T$ instead of $\tilde C_T$ as in \cite{Hung:2011nu}. The two are related by $C_T = \frac{d+1}{d-1}\frac{\Gamma(d+1)}{\pi^d} \tilde C_T$.}
\begin{equation}
\begin{split}
a_d^* &= \frac{\pi^{d/2}}{\Gamma(d/2)}\left(\frac{\tilde L}{l_P} \right)^{d-1} \left[1-2\frac{d-1}{d-3}\lambda g_{\infty} \right]\,,
\\
C_T & = \frac{\Gamma(d+2)}{\pi^{d/2}(d-1)\Gamma(d/2)}\left(\frac{\tilde L}{l_P} \right)^{d-1} \left[1-2\lambda g_{\infty} \right]\,.
\end{split}
\end{equation}
Again, $C_T$ is the central charge appearing in the vacuum two-point correlator \reef{ozzy} of the stress tensor, while $a_d^*$ is the universal coefficient appearing in the entanglement entropy of a sphere in the CFT vacuum \cite{Myers:2010xs,Myers:2010tj,Casini:2011kv}.

Now as in the Einstein analysis, we express $C_D$ for GB gravity as a function of the integration constant $\be_n$,
\begin{equation}\label{CDGB}
\frac{C_D}{n}  =  \frac{d\,\Gamma(d+1) }{(d-1)\pi^{d/2-2}\Gamma({d}/{2})}
\left(
\sqrt{1-4 \lambda }\,(d-2)\left(\frac{\tilde L}{\lp}\right)^{d-1}   (\beta_n-\beta_1)
+ \frac{h_n}{2\pi n}\right) \,.
\end{equation}

\begin{figure}
        \centering
        \begin{subfigure}[b]{0.45\textwidth}
                \includegraphics[width=\textwidth]{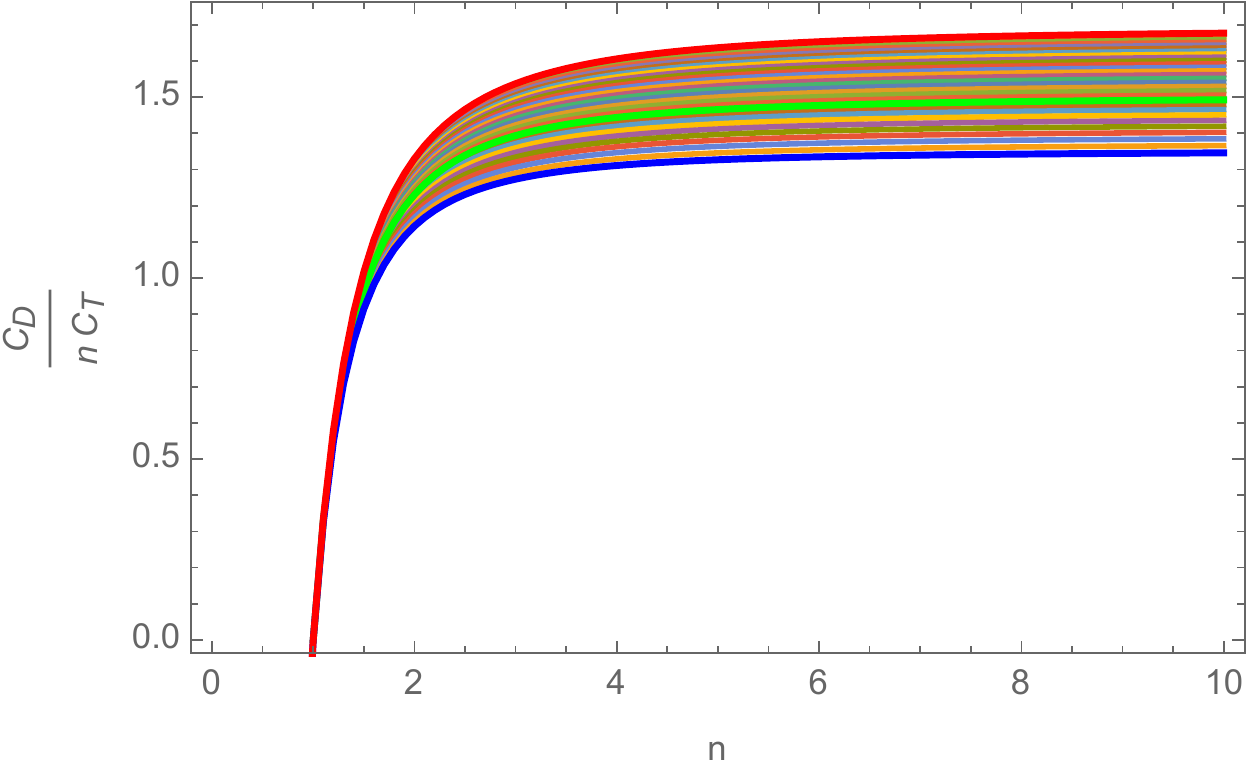}
                \caption{$d=4$}
                \label{fig1:frb}
        \end{subfigure}
~\quad
        \begin{subfigure}[b]{0.45\textwidth}
                \includegraphics[width=\textwidth]{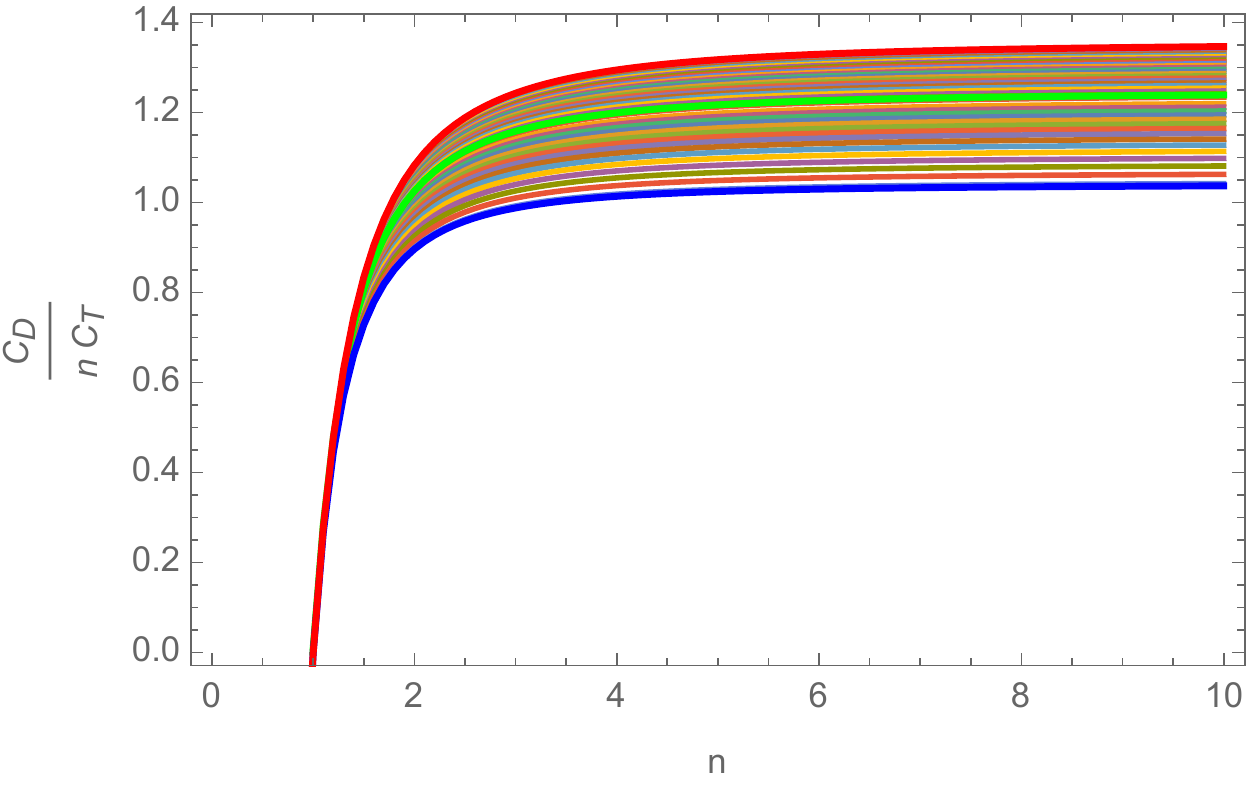}
                \caption{$d=5$}
                \label{fig1:frc}
        \end{subfigure}
        \caption{$C_D/(n C_T)$ as a function of $n$ for $d=4$ and $d=5$ and for different values of Gauss-Bonnet coupling between the unitarity bounds given in eq.~\eqref{bounds}. The red curve gives the negative lower bound while the blue line corresponds to the positive upper bound. Highlighted in green is the Einstein gravity solution (\ie $\la=0$) that of course agrees with solutions found in the previous section. Intermediate curves correspond to intermediate values of the coupling in steps of $\Delta \lambda = 0.01$.  }
        \label{fig_cd_gb}
\end{figure}

\begin{figure}
        \centering
        \begin{subfigure}[b]{0.45\textwidth}
                \includegraphics[width=\textwidth]{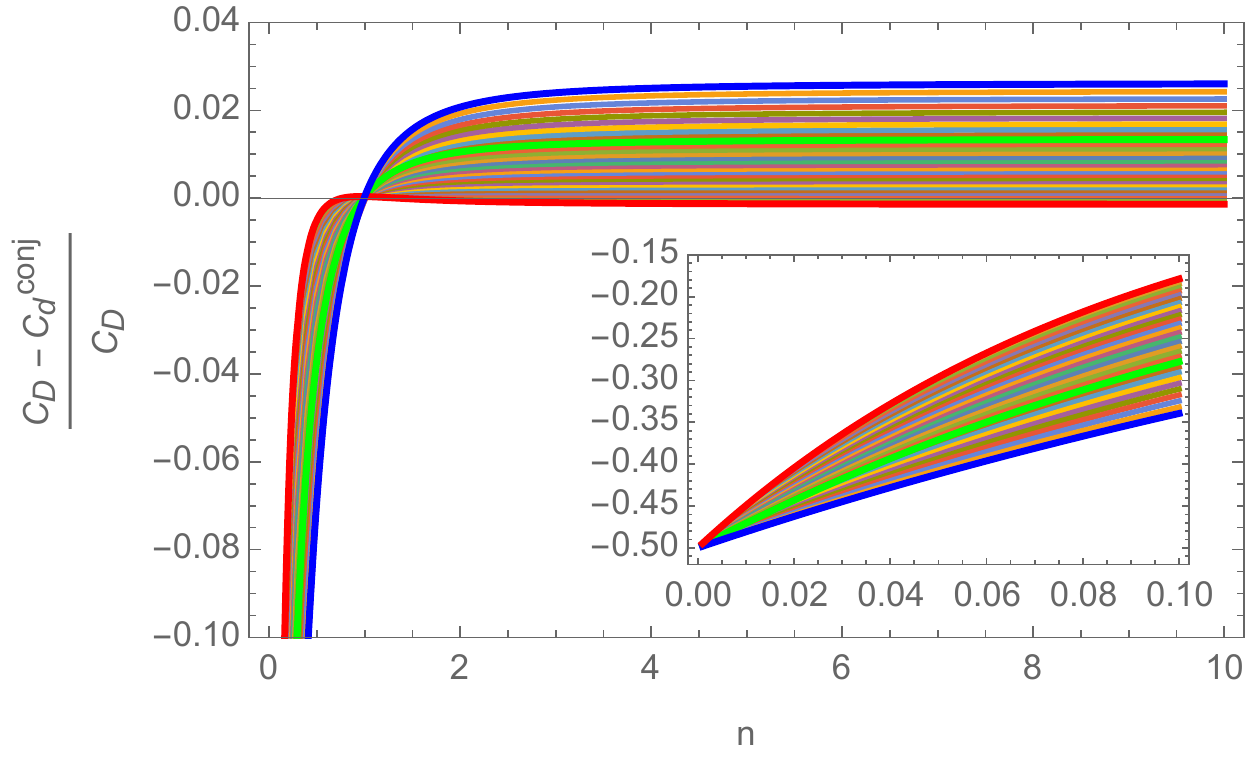}
                \caption{$d=4$}
                \label{fig2:frb}
        \end{subfigure}
~\quad
        \begin{subfigure}[b]{0.45\textwidth}
                \includegraphics[width=\textwidth]{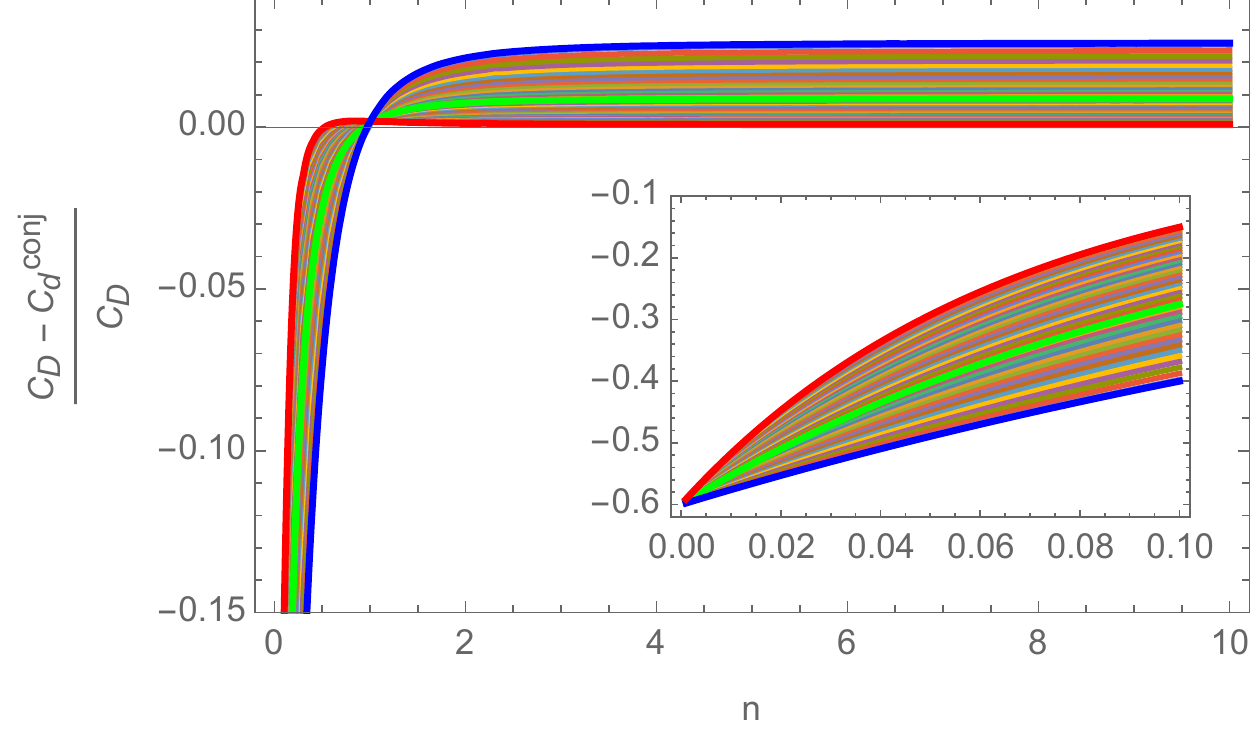}
                \caption{$d=5$}
                \label{fig2:frc}
        \end{subfigure}
        \caption{The relative error $(C_D-C_D^{\conj})/C_D$ as a function of $n$ for $d=4$ and $d=5$ and for different values of GB coupling between the unitarity bounds given in eq.~\eqref{bounds}. The red curve gives the lower bound while the blue line corresponds to the upper bound. Highlighted in green is the Einstein gravity solution (\ie $\la=0$) that of course agrees with solutions found in the previous section. Intermediate curves correspond to intermediate values of $\la$ in steps of $\Delta \lambda = 0.01$. In the inset, we present the solutions near $n=0$ and show that independently of the GB coupling the curves approach the Einstein gravity value $(2-d)/d$. }
        \label{fig_cd_rel}
\end{figure}

We solve numerically for $\beta_n$ in $d=4$ and $d=5$ and the results for $C_D$ are shown in figs.~\ref{fig_cd_gb} and \ref{fig_cd_rel}. The curve corresponding to Einstein gravity (\ie $\la=0$) is highlighted in green. One can immediately see that, on one hand, the qualitative behavior of $C_D$ in Einstein gravity is shared by all the curves for Gauss-Bonnet gravity. On the other hand, by tuning the coupling $\lambda$ one can substantially reduce the discrepancy between $C_D$ and $C_D^{\conj}$ when $n>1$. In particular, in $d=4$, if we choose $\la$ at the lower unitarity bound, $C_D-C_D^{\conj}$ becomes negative for $n$ sufficiently large. Since the relative error is asymptotically constant, this implies that there is an allowed value of the coupling for which the conjecture \eqref{conj} is fulfilled at large $n$. However, as one might have expected, there is no value of $\la$ for which the conjecture is satisfied for all values of the \ren\ index.

Now we turn into the perturbative expansion close to $n=1$, that as in the Einstein case admits an analytic treatment.
Near $n=1$, we can write $\beta_n$ as
\begin{equation}
\begin{split}
\beta_n^{(d=4)} &= -\frac{1}{8}+\frac{n-1}{12}-\left(\frac{17}{144}+\frac{1}{27 \sqrt{1-4 \lambda }}\right) (n-1)^2+\ldots,
\\
\beta_n^{(d=5)} & =  \frac{1}{20} (n-1) - \left(\frac{731}{9600}+\frac{190}{9600\sqrt{1-4\lambda}}\right) (n-1)^2 +\dots,
\\
\beta_n^{(d=6)} & =  -\frac{1}{16}+\frac{n-1}{30}
-\left(\frac{241}{4500}+\frac{51}{4500\sqrt{1-4 \lambda}}\right) (n-1)^2+\dots.
\end{split}
\end{equation}
This result, together with eqs.~\eqref{hnGB} and \eqref{CDGB}, yields\footnote{\rev{We would like to thank Rong-Xin Miao for pointing out a typo in these formulas for $d=6$ in a previous version of the paper.}}
\begin{equation}
\begin{split}
\frac{C_D^{(d=4)}}{C_T}&= \frac{2\pi^2}{5}  (n-1) -\frac{\pi ^2}{30}  \left(3+\frac{8}{\sqrt{1-4 \lambda }}\right) (n-1)^2+ O(n-1)^3,
\\
\frac{C_D^{(d=5)}}{C_T}&= \frac{\pi ^2}{3}  (n-1) - \frac{\pi ^2}{60}  \left(9+\frac{10}{\sqrt{1-4 \lambda }}\right) (n-1)^2+ O(n-1)^3,
\\
\frac{C_D^{(d=6)}}{C_T}& = \frac{2}{7} \pi ^2 (n-1)-\frac{\pi ^2}{105 }\frac{\left(17 \sqrt{1-4 \lambda}+12\right) }{\sqrt{1-4 \lambda}}(n-1)^2
+ O(n-1)^3\,.
\end{split}
\end{equation}
With the conjectured expression \reef{conj} for $C_D$, we find that in GB gravity
\begin{equation}
\begin{split}
\frac{C_D^{\conj}{}^{(d=4)}}{C_T}& = \frac{2\pi^2}{5}  (n-1)-\frac{\pi ^2}{45}  \left(3+\frac{14}{\sqrt{1-4 \lambda }}\right) (n-1)^2 + O(n-1)^3,
\\
\frac{C_D^{\conj}{}^{(d=5)}}{C_T}& = \frac{\pi ^2}{3}  (n-1)-\frac{\pi ^2}{96}  \left(13+\frac{18}{\sqrt{1-4 \lambda }}\right) (n-1)^2 + O(n-1)^3,
\\
\frac{C_D^{\conj}{}^{(d=6)}}{C_T} &= \frac{2}{7} \pi ^2 (n-1)-\frac{\pi ^2 }{175}\frac{\left(27 \sqrt{1-4 \lambda}+22\right) }{ \sqrt{1-4 \lambda}}(n-1)^2+ O(n-1)^3.
\end{split}
\end{equation}
Hence we again recover the necessary agreement at linear order in $(n-1)$, but the expressions differ at the quadratic order. In particular, the relative mismatch, which now depends on the GB coupling, becomes
\begin{equation}
\begin{split}
\frac{C_D^{(d=4)}-C_D^{\conj}{}^{(d=4)}}{C_D^{(d=4)}} = \left(\frac{1}{9 \sqrt{1-4 \lambda}}-\frac{1}{12}\right) (n-1)+O\left(n-1\right)^2,
\\
\frac{C_D^{(d=5)}-C_D^{\conj}{}^{(d=5)}}{C_D^{(d=5)}} = \left(\frac{1}{16 \sqrt{1-4 \lambda}}-\frac{7}{160}\right) (n-1)+O(n-1)^2,
\\
\frac{C_D^{(d=6)}-C_D^{\conj}{}^{(d=6)}}{C_D^{(d=6)}} = \frac{1}{75} \left(\frac{3}{\sqrt{1-4 \lambda}}-2\right) (n-1)+O\left(n-1\right)^2.
\end{split}
\end{equation}
When $\lambda=0$, we correctly reproduce the results of Einstein gravity. However, we now see that the coupling can be tuned to eliminate the discrepancy at the next order as well. It turns out that the value of $\la$  required  to produce agreement with eq.~\eqref{conj} at order $(n-1)^2$ can be expressed as
\begin{equation}
\lambda_{min} = -\frac{(3 d+2) (d-2)}{4 (d+2)^2}\,,
\end{equation}
for $d=4,5$ and 6. Surprisingly, this value corresponds precisely to the lower bound in eq.~\eqref{bounds} arising from unitarity constraints. We discuss possible implications of this observation in section \ref{sec:discussion}.

Finally, we would like to consider the limit $n\to0$, which is also amenable to an analytic understanding. For any value of $\lambda$, we find again that $\beta_n\sim 1/n^{d-1}$, while $x_n \sim 1/n$. Hence, $\beta_n$ can be neglected in eq.~\reef{CDGB} at leading order in $1/n$, and we obtain
\begin{equation}
\frac{C_D}{C_T}=-\left( \frac{1}{dn} \right)^{d-1}
\left(
\frac{\pi ^2 \left(\sqrt{1-4 \lambda }+1\right) \left(\frac{1-\sqrt{1-4 \lambda }}{\lambda }\right)^d}{4 (d+1) \sqrt{1-4 \lambda }}+  O(n) \right),
\end{equation}
\begin{equation}
\frac{C_D^{\conj}}{C_T}=-\left( \frac{1}{dn} \right)^{d-1}
\left(\frac{ \pi ^2(d-1)  \left(\sqrt{1-4 \lambda}+1\right) \left(\frac{1-\sqrt{1-4 \lambda}}{ \lambda}\right)^{d}}{2 d(d+1) \sqrt{1-4 \lambda}}+O\left(n\right)\right).
\end{equation}
In particular then, we find
\begin{equation}
 \frac{C_D-C_D^{\conj}}{C_D}  = - \frac{d-2}{d} + O(n)\,,
 \label{nto0GB}
\end{equation}
which is remarkably independent of the coupling, and therefore equal to the result \reef{rel_near0} found for Einstein gravity. The convergence of the curves corresponding to different values of the GB coupling to the ratio \eqref{nto0GB} is plotted in the inset of figure \ref{fig_cd_rel}. By comparison with eq.~\eqref{CDGB}, we notice that the universality of the small $n$ limit is exclusively due to the scaling of $\beta_n$ with respect to the conformal weight $h_n$. The behavior of $h_n$ is fixed by simple thermodynamic considerations, which are reliable in the high temperature limit ($n\to0$), where all other scales are negligible. It would be tempting to look for a similar argument, which could predict the scaling of $\beta_n$ as well. However, the existence of theories that satisfy the conjecture at every value of $n$ implies that the two statements cannot have the same degree of universality.

\section{Discussion}
\label{sec:discussion}

The results of this paper confirm and extend the qualitative picture which emerged after the four-dimensional study \cite{Dong:2016wcf}. Twist operators in strongly interacting holographic CFTs do not obey the conjectured relation $C_D=C_D^{\conj}$. However, the relation is only mildly violated for a large range of values of $n$. It would be useful to understand whether this is accidental or not. One way of tackling this question would be to study the stability of this qualitative picture under further higher derivative corrections. More generally, our understanding about the conjectured relation $C_D=C_D^{\conj}$ seems incomplete. Although the relation is obeyed by some examples of free theories in $d=3$ \cite{Dowker1,Dowker2} and $d=4$ \cite{Bianchi:2015liz}, it has not been established whether its violation is only a consequence of the presence of interactions. Further investigations in the context of free theories will be required in order to answer this question. More generally, we know that when the conjecture \eqref{conj} is satisfied, the singularities of the defect OPE with the stress tensor are simplified \cite{Bianchi:2015liz}, but we do not yet have an understanding of the consequences of this fact on the structure of entanglement. Finally, we note that this problem can be reformulated more broadly: what are the properties required for a defect to obey the conjecture \eqref{conj}? In fact, this issue has been a question of interest in the context of gauge theories as well. A similar relation between the Bremsstrahlung function -- which is the analogue of $C_D$ in that context -- and the conformal weight is obeyed by a class of Wilson lines. However, the theories in which this happens have not been classified yet \cite{Lewkowycz:2013laa,Fiol:2015spa}. In fact, in $d=3$ the relation conjectured in \cite{Lewkowycz:2013laa} reduces to our relation $C_D=C_D^{\conj}$.

Beyond the broad picture, the results of our work raise more detailed points of discussion. We have found that the $n\to0$ limit of the ratio $C_D/C_D^{\conj}$ is not affected by the presence of the Gauss-Bonnet coupling in the bulk theory. However it is certainly different in the free CFTs where eq.~\reef{conj} has been proven to be satisfied. It would be interesting to check whether this limit has the same value for other higher curvature theories of gravity. If true, the small $n$ limit would exhibit an insensitivity to higher derivative corrections which is not shared by many other quantities. The prototypical example of this behavior is the Lyapunov exponent in holographic theories \cite{Maldacena:2015waa}.

In general, our holographic result \reef{CDGB} for $C_D$ with Gauss-Bonnet gravity depends on the Gauss-Bonnet coupling and so we may consider tuning $\la$ to achieve $C_D=C_D^{\conj}$. While the above result for $n=0$ shows that we can not achieve this equality for all values of $n$, it is still possible to achieve this equality at special values of $n$. In particular, in the limit $n\to \infty$, we found numerically that in four dimensions there is a value of $\la$ (within the unitarity bounds \reef{bounds}) for which the conjecture is asymptotically satisfied -- see the comment below eq.~\eqref{CDGB}. Similarly close to $n=1$, the analytic solutions allowed us to make a more interesting statement. Of course, the results \cite{Faulkner:2015csl} demand agreement of $\partial_nC_D|_{n=1}$ with the conjectured value for any value of the GB coupling. However, tuning  $\la$ allows the conjecture to be fulfilled up to order $(n-1)^2$. The surprising result, however, is that the particular value of $\la$ which achieves this tuning precisely saturates the lower unitarity bound given in eq.~\reef{bounds} for $d=4$, 
5 and 6.\footnote{\rev{Shortly after the appearance of our paper, ref.~\cite{Chu:2016tps} appeared in which it was shown that the relation between the fulfillment of the conjecture at order $(n-1)^2$ and the lower unitarity bound is a universal property of general higher curvature gravity theories in general dimensions. The authors of \cite{Chu:2016tps} also wrote down explicitly the relation between $\partial^2_n C_D|_{n=1}$, $\partial^2_n h_n|_{n=1}$ and the coefficients in the three-point function of the stress tensor which we describe in the following paragraph and suggested a universal law which is obeyed by free scalars, free fermions, free conformal tensor fields and
CFTs with holographic dual.}}

The latter observation is rather interesting and deserves further comment. The result would perhaps be less surprising if one could prove that $\partial^2_n C_D|_{n=1}$ is only sensitive to a restricted set of CFT data. This suggestion might be motivated by the results for $\partial_nC_D|_{n=1}$ \cite{Faulkner:2015csl} and for the first two derivatives of $h_n$ \cite{Hung:2014npa, Lee:2014zaa}. The intuition is inspired by the fact that derivatives with respect to the \ren\ index are equivalent to repeated insertions of the modular Hamiltonian in the partition function. If applied to the one-point function \eqref{Tgoodframe}, this would fix both $\partial^2_nh_n|_{n=1}$ and $\partial^2_nC_D|_{n=1}$ as a linear combination of the three coefficients appearing in the three-point function of the stress-tensor, which in general dimensions are often denoted $\mc{A},\,\mc{B},\,\mc{C}$ \cite{Osborn:1993cr,Erdmenger:1996yc}. While this argument is correct in the case of $h_n$ \cite{Hung:2014npa}, when dealing with shape deformations further subtleties arise, and we will study them elsewhere.
If the argument given above was correct, the fulfillment of the conjecture at order $(n-1)^2$ would be a general property of theories that saturate the Hofman-Maldacena bound \cite{Hofman:2008ar}, from which the unitarity bound for Gauss-Bonnet gravity was derived \cite{Buchel:2009sk}. From our results for Gauss-Bonnet gravity, it is not difficult to derive  a simple ansatz for $\partial^2_nC_D|_{n=1}$, up to one free coefficient. In $d=4$, also the last coefficient can be fixed thanks to the fact that free scalars satisfy the conjecture \eqref{conj}.
However, instead of writing the explicit linear combination of $\mc{A},\,\mc{B}$ and $\mc{C}$, let us point out that this proposal meets an obstruction in $d=3$.
In this case, we know that free scalars and fermions obey the conjecture, while the holographic dual of Einstein gravity does not. At the same time, in $d=3$ there are only two independent parameters in the three-point function of the stress-tensor. This leads to a system with three equations, \ie for scalars, fermions and Einstein gravity, and two unknowns. It is easy to verify that the system does not have a solution. This proves that in three dimensions $\partial^2_nC_D|_{n=1}$ is not determined by the three-point function of the stress tensor, or at least not in a way which is linear in the parameters. Although it seems unlikely that $d=3$ plays a special role, rejecting this proposal in general
would leave us without an answer to the original question: the special role played by the unitarity bound of the Gauss-Bonnet coupling would remain mysterious. Certainly, these observations imply that it would be interesting to attempt a first principle computation of $\partial^2_nC_D|_{n=1}$. On the other hand, even one single further example in four dimensions would provide a non-trivial check. One could either perform one more free field computation \cite{future}
or study another holographic example along the lines of the present paper.

It is interesting to ask under which conditions the response of the holographic twist operator to small deformations might change qualitatively with respect to the results of this work. For example, `phase transitions' in the \ren\ entropy have been observed to take place at a critical value of the \ren\ index in certain theories \cite{max}. This behaviour has also been observed in a holographic setting for spherical regions \cite{Belin:2013dva,pufu}. This happens when the defect CFT has a sufficiently low-dimensional scalar operator, in which case the dual hyperbolic black hole solution would be unstable towards the development of scalar hair at low temperatures (large $n$). We expect that in such theories the \ren\ entropy for non-spherical regions should have similar phase transitions. It would be interesting to study what effects the phase transitions have on $C_D$ in the presence of such low-dimensional scalar operators.

Finally, let us emphasize that the strategy that we have employed in this paper can easily be adapted to the study of deformations of any holographic defect. Indeed, the information about the \ren\ twist operator enters the computation only through the choice of the gravitational background. In particular, the fact that this approach only requires linear order perturbation theory may prove useful in many other situations.

\section*{Acknowledgments}
We would like to thank Misha Smolkin for valuable collaboration at an early stage of this project.
DAG would like to thank the organizers and participants of the “YKIS 2016: Quantum Matter, Spacetime and Information” conference held at YITP, Kyoto between June 13-17 where the results of this paper were presented for the first time. SC, LB and MM would like to thank the organizers of the GGI workshop ``Conformal Field Theories and Renormalization Group Flows in Dimensions $d>2$'' for hospitality and for giving the opportunity to SC to give a talk on the results of this work on June 30.
MM would like to thank Davide Gaiotto for useful discussions. Research at Perimeter Institute is supported by the Government of Canada through Industry Canada and by the Province of Ontario through the Ministry of Research \& Innovation. The work of LB is supported by Deutsche Forschungsgemeinschaft in Sonderforschungsbereich 676 ``Particles, Strings, and the Early Universe''.
SC acknowledges support from an Israeli Women in Science  Fellowship from the Israeli Council of Higher Education.
XD is supported in part by the Department of Energy under Grant No.\ DE-SC0009988 and by a Zurich Financial Services Membership at the Institute for Advanced Study.
XD would also like to thank the Perimeter Institute and the ``It from Qubit" summer school for hosting visits at various stages of this collaboration. RCM is supported by funding from the Natural Sciences and Engineering Research Council of Canada, from the Canadian Institute for Advanced Research and from the Simons Foundation through the ``It from Qubit'' collaboration.

\appendix

\section{Expanding Two-Point Functions as Distributions}
\label{deltas}
In this appendix we provide the details needed to derive equations \eqref{Tdeltas_a_ij}-\eqref{Tdeltas_a_bc}. As mentioned in the main text the singular terms in the short distance expansion $|x|\rightarrow 0$ of the two-point function $\langle D_a(w) T_{\mu\nu}(z) \rangle$ can be written in the weak limit in terms of delta functions in the $d-2$ parallel directions with support at $w=0$, where we recall that $z^\m=(x,y)$, and we further fixed $y=0$. Keep in mind that the expressions that we write in this appendix only hold inside integrals when multiplied by a test function that decays fast enough at infinity and is regular at zero.

The general expansion \emph{up to terms which are regular as $|x|\rightarrow 0 $} reads
\begin{equation}\label{eq:kernel_id}
\begin{split}
&\frac{w^{2     \alpha}}{(x^2+w^2)^{d-1+\beta}}
=  \,\frac{\pi^\frac{d-2}{2}}{\Gamma(\frac{d-2}{2})\Gamma(d+\beta-1)}\times
\\
& ~~~~~~~~
\sum_{n=0\atop n\,\text{even}}^{{}_{d+2(\beta-\alpha)-1}} (\pa^2)^{\frac{n}{2}}\delta^{d-2}(w)\left[\frac{(n-1)!!(d-4)!!}{n!(d-4+n)!!}
\right]
\frac{\Gamma(\frac{d-n}{2}-\alpha+\beta) \Gamma(\frac{d+n}{2}+\alpha-1)}{|x|^{d+2\beta-2 \alpha-n}}
\end{split}
\end{equation}
Similar formulas for tensorial structures of $w$ can be derived by differentiating identities of the form \eqref{eq:kernel_id}.
In the rest of this appendix we present a derivation of the formula \eqref{eq:kernel_id} as well as a list of useful identities that can be deduced from it.

\subsection{Derivation of the Kernel Formula}
Consider the kernel
\begin{equation}
K(w,x) = \frac{w^{2\alpha}}{(x^2+w^2)^{d-1+\beta}}
\end{equation}
and a test function $f(w)$ which is smooth at $w=0$ and decays
strong enough (such that \eqref{eq:kernelInt} converges)
when $w\rightarrow\infty$
and define the integral
\begin{equation}\label{eq:kernelInt}
I(x) = \int d^{d-2}w f(w) K(w,x).
\end{equation}
We can split the integration domain between $|w|\leq 1$ and $|w|>1$. The exterior region is convergent when $|x|\rightarrow 0$ and therefore does not contribute to the divergent terms in \eqref{eq:kernel_id}. The function $f(w)$ can be Taylor expanded in the inside domain as follows:
\begin{equation}
I(x) = \sum_{n=0}^{\infty} \frac{1}{n!} \del_{i_1} \ldots \del_{i_n} f(0)       \int_{|w|<1} d^{d-2} w \, w^{i_1} \ldots w^{i_n} K(w,x) + \text{regular},
\end{equation}
where ``regular'' stands for terms which are regular at $x \rightarrow 0$. Using a change of variables $w^i = y^i |x|$ and symmetry considerations on the tensor structure inside the integral (which after integration can only depend on Kronecker delta functions) we obtain
\begin{equation}
\begin{split}
I(x) = & \, \sum_{n=0}^{\infty} \frac{1}{n!} \del_{i_1} \ldots \del_{i_n} f(0)    \, \frac{1}{|x|^{d-n+2\beta-2\alpha}}\int_{|y|<1/|x|} d^{d-2} y \, \frac{ y^{2\alpha} y^{i_1} \ldots y^{i_n}}{(1+y^2)^{d-1+\beta}} + \text{regular}
\\
= & \, \sum_{n=0}^{\infty} \frac{1}{n!} \del_{i_1} \ldots \del_{i_n} f(0)       \, \frac{1}{|x|^{d-n+2\beta-2\alpha}}
\frac{\delta^{i_1i_2} \ldots \delta^{i_{n-1}i_n}+\text{permutations}}{\text{normalization}}\times
\\
&
~~~~~~~~~~~~~~~~~~~~~~~~~\times\Omega_{d-3} \int_{|y|<1/|x|} d y \, \frac{y^{n+d-3+2\alpha}}{(1+y^2)^{d-1+\beta}} + \text{regular}
\\
= & \, \sum_{n=0}^{\infty} \frac{1}{n!} \del_{i_1} \ldots \del_{i_n} f(0)       \,
\frac{\delta^{i_1i_2} \ldots \delta^{i_{n-1}i_n}+\text{permutations}}{\text{normalization}}\times
\\
&
~~~~~~~~~~~~~\times\frac{1}{|x|^{d-n+2\beta-2\alpha}} \frac{2 \pi^{\frac{d-2}{2}}}{\Gamma(\frac{d-2}{2})}
\frac{\Gamma(\frac{d - n}{2}-\alpha+\beta) \Gamma(\frac{d + n}{2}+\alpha-1)}{2 \Gamma(d+\beta-1)}
+ \text{regular}.
\end{split}
\end{equation}
In the weak limit we replace $f(0)$ by $\delta^{d-2}(w)$ to express the kernel $K(w,x)$ outside the integral as
\begin{equation}
\begin{split}
\frac{w^{2      \alpha}}{(x^2+w^2)^{d-1+\beta}} =  \frac{\pi^\frac{d-2}{2}}{\Gamma(\frac{d-2}{2})\Gamma(d+\beta-1)}
\sum_{n=0}^{\infty} &\frac{P_n \, \Gamma(\frac{d-n}{2}-\alpha+\beta) \Gamma(\frac{d+n}{2}+\alpha-1)}{|x|^{d+2\beta-2 \alpha-n}}
\end{split}
\end{equation}
where
\begin{equation}
P_n = \frac{1}{n!} \del_{i_1} \ldots \del_{i_n} \delta^{d-2}(w)
\left[ \frac{\delta^{i_1 i_2} \ldots \delta^{i_{n-1} i_n} + \text{permutations}}{\text{normalization}} \right],
\end{equation}
and the normalization is chosen such that the term in the square bracket traces to one. Combinatorial  arguments then lead to the simplified form \eqref{eq:kernel_id}.

\subsection{List of Formulas for Kernels}

We use \eqref{eq:kernel_id} to derive the following:
\begin{align}\label{d-1kernel}
\begin{split}
& \frac{1}{(x^2+w^2)^{d-1}} =  \frac{\pi^{\frac{d-2}{2}} \Gamma(\frac{d}{2})}{  \Gamma(d-1)} \left(\frac{\delta^{d-2}(w)}{|x|^{d}} + \frac{\del^2 \delta^{d-2}(w)}{2(d-2)|x|^{d-2}}
 \right)
+ \ldots,
\end{split}\\
\begin{split}\label{dkernel}
& \frac{1}{(x^2+w^2)^{d}} =  \frac{\pi^{\frac{d-2}{2}} \Gamma(\frac{d}{2})}{ 2 \Gamma(d)} \left(\frac{d \,\delta^{d-2}(w)}{ \,|x|^{d+2}} + \frac{\del^2 \delta^{d-2}(w)}{2|x|^{d}}
 \right)
+ \ldots
,
\end{split}\\
\begin{split}\label{dp1kernel}
& \frac{1}{(x^2+w^2)^{d+1}} =
\frac{ \pi^{\frac{d-2}{2}} \Gamma(\frac{d }{2})}{ 4\Gamma(d)}
\left[\frac{\delta^{d-2}(w)}{|x|^{d+4}}
(d+2)
+
 \frac{\del^2\delta^{d-2}(w)}{2 |x|^{d+2}}
\right]
+ \ldots ,
\end{split}\\
\begin{split}\label{w2dkernel}
& \frac{w^2}{(x^2+w^2)^{d}} =  \frac{ \pi^{\frac{d-2}{2}} \Gamma(\frac{d }{2})}{ 2\Gamma(d)} \left[\frac{\delta^{d-2}(w)}{|x|^{d}}
(d-2)
+
 \frac{d}{2(d-2)}\frac{\del^2\delta^{d-2}(w)}{|x|^{d-2}}
\right]
+ \ldots ,
\end{split}\\
\begin{split}\label{w2dp1kernel}
& \frac{w^2}{(x^2+w^2)^{d+1}} =  \frac{ \pi^{\frac{d-2}{2}} \Gamma(\frac{d }{2})}{ 4\Gamma(d)}
\left[\frac{\delta^{d-2}(w)}{|x|^{d+2}}
(d-2)
+
 \frac{\del^2\delta^{d-2}(w)}{2|x|^{d}}
\right]
+ \ldots ,
\end{split}\\
\begin{split}\label{w4dp1kernel}
& \frac{w^4}{(x^2+w^2)^{d+1}} =  \frac{ \pi^{\frac{d-2}{2}} \Gamma(\frac{d }{2})}{ 4\Gamma(d)} \left[\frac{\delta^{d-2}(w)}{|x|^{d}}
(d-2)
+
 \frac{d+2}{2(d-2)}\frac{\del^2\delta^{d-2}(w)}{|x|^{d-2}}
\right]
+ \ldots .
\end{split}
\end{align}
Differentiating \eqref{eq:kernel_id} with respect to $w_i$ we get
\begin{equation}
\frac{w^{2      \alpha}w_i }{(x^2+w^2)^{d+\beta}} =
\frac{1}{2(1-d-\beta)} \left[\del_i \left(\frac{w^{2    \alpha}}{(x^2+w^2)^{d-1+\beta}} \right) -\frac{2\alpha w^{2     (\alpha-1)}w_i}{(x^2+w^2)^{d-1+\beta}} \right].
\end{equation}
We can then use \eqref{d-1kernel} and \eqref{dkernel} to show
\begin{equation}\label{widkernel}
\frac{ w^i}{(x^2+w^2)^d}=-  \frac{\pi^{\frac{d-2}{2}} \Gamma(\frac{d}{2})}{ 2 \Gamma(d)} \left(\frac{\del^i\delta^{d-2}(w)}{|x|^{d}} + \frac{\del^i\del^2 \delta^{d-2}(w)}{2(d-2)|x|^{d-2}}
 \right)
+ \ldots
,
\end{equation}
and
\begin{equation}
\frac{w^i}{(x^2+w^2)^{d+1}} = - \frac{\pi^{\frac{d-2}{2}} \Gamma(\frac{d}{2})}{ 4 \Gamma(d+1)} \left(\frac{d \,\del^i\delta^{d-2}(w)}{ \,|x|^{d+2}} + \frac{\del^i\del^2 \delta^{d-2}(w)}{2|x|^{d}}
 \right)
+ \ldots
,
\end{equation}
and \eqref{w2dkernel} and \eqref{widkernel} to get
\begin{equation}
\frac{w^2 w^i}{(x^2+w^2)^{d+1}}
 =  -  \frac{\pi^{\frac{d-2}{2}} \Gamma(\frac{d}{2})}{ 4 \Gamma(d+1)}\left[
 \frac{d\, \del^i\delta^{d-2}(w)}{|x|^{d}}
+
\frac{(d+2)}{2(d-2)}
 \frac{ \del^i\del^2\delta^{d-2}(w)}{|x|^{d-2}}
\right]
+ \ldots
.
\end{equation}
Differentiating \eqref{d-1kernel} twice and using \eqref{dkernel} we can also show
\begin{equation}
\frac{ \, w^i w^j}{(x^2+w^2)^{d+1}} = \frac{\Gamma(\frac{d}{2}) \pi^{\frac{d-2}{2}}}{4\Gamma(d+1)} \left[\delta^{ij} \left(\frac{d \, \delta^{d-2}(w)}{|x|^{d+2}} +\frac{\del^2 \delta^{d-2}(w)}{2 |x|^d}
\right) +\frac{\del^i \del^j \delta^{d-2}(w) }{|x|^d} \right].
\end{equation}
This concludes the ingredients needed for the derivation of equations \eqref{Tdeltas_a_ij}-\eqref{Tdeltas_a_bc}.

\section{Details of Holographic Renormalization for Einstein Gravity}
\label{app_holo_ren}
In this appendix we give more details on how to obtain the expectation value of the stress tensor in Einstein gravity.
The procedure for GB gravity is similar, with the difference that the final expression for the expectation value changes to eq. (\ref{gorp}) \cite{Faulkner:2013ica}. 
As described in section \ref{holgra}, the expectation value of the stress tensor in the deformed geometry can be computed using eqs.~\eqref{FGmetric}--\eqref{holT}. 
As noted in the main text, it is not necessary for our purposes to compute explicitly $\mathcal X$ since all $h_{(m)}$ with $m<d$ are independent of $x_n$ (equivalently of the \ren\ index $n$) and then, they will cancel after the subtraction in \eqref{eq:CFToutputkn}. We keep track of the anomalous contributions for completeness.

In order to write the metric \eqref{metric} in the FG coordinates, as in equation \eqref{FGmetric}, we work order by order in an expansion in $z$ (that we define perturbatively below). We reach the desired coordinate system by two successive changes of coordinates. First we define $r \equiv L^2/\tilde{z}$ and then we redefine $\tilde{z} = z (1+ c_1 z^2 +\ldots+ c_d z^d+\ldots)$. Using the blackening factor \eqref{eq:black_factor} we can fix the constants $c_i$ such that
\begin{eqnarray}
\frac{dr^2}{g(r)} = \frac{dz^2 (\partial \tilde{z}/\partial z)^2}{\tilde z^4 g(r(z))} = \frac{L^2}{z^2} dz^2 (1+O(z^{d+1})) \,.
\end{eqnarray}
We obtain
\begin{equation}
\begin{split}\label{eq:z_zt_expansion}
d  = & \, 3\quad , \quad \tilde
z =z \left(1-\frac{z^2}{4 L^2}-\frac{x_n \left(x_n^2-1\right)}{6} \frac{z^3}{L^3}+\ldots\right) ,  \\
d = & \, 4 \quad , \quad
\tilde z =z \left(1-\frac{z^2}{4 L^2 }-\frac{x_n^2( x_n^2-1)-1/2}{8}  \frac{z^4}{L^4} +\ldots \right), \\
d  = & 5 \quad , \quad \tilde z =z \left(1-\frac{z^2}{4 L^2}+\frac{z^4}{16 L^4}-\frac{x_n^3\left(x_n^2-1\right)}{10}  \frac{z^5}{L^5}+\ldots\right) ,
\\
d  = & 6 \quad , \quad
\tilde z =z \left(1-\frac{z^2}{4 L^2}+\frac{z^4}{16 L^4}   -\frac{x_n^4(x_n^2- 1)+3/16}{12}  \frac{z^6}{L^6} +\ldots\right) .
\end{split}
\end{equation}
Note that the $x_n$ (equivalently $n$) dependence starts appearing only at order $z^d$. This is due to the form of the blackening factor \eqref{eq:black_factor} in which the dependence on $x_n$ starts at the $d$-th subleading order in the boundary expansion. It is then obvious that only $h_{(d)\mu \nu}$ of \eqref{smallz} will depend on $n$ while the lower order $h_{(m)}$'s entering in the anomalous functional $\mathcal{X}$ will not.
In order to get our FG metric we now need to expand the components of the original metric \eqref{metric} in powers of $z$. Equivalently, in terms of the $h_{\mu\nu}$ of equation \eqref{FGmetric} we need to expand the factors in square brackets in the following expression:
\begin{equation}
\begin{split}
h_{\mu\nu} dx^\mu dx^\nu
=  &
\left[\frac{z^2}{L^2} g(r(z))\right] d\tau^2
+ \left[\left(\frac{z}{\tilde z}\right)^2 \right]\frac{d\rho^2}{\rho^2}
\\&
+ \left[\frac{4}{d-2}  \left(\frac{z}{\tilde z}\right)^2 v(r(z))\right]  \left(\partial_i \, K^b x_b\right) \frac{d\rho }{\rho}  dy^i
\\
& + \frac{1}{\rho^2}   \left(\left[\left(\frac{z}{\tilde z}\right)^2 \right] \delta_{ij}
 +\left[2 \, \left(\frac{z}{\tilde z}\right)^2 k(r(z))\right] \tilde{K}^a_{ij}x_a  \right) dy^i dy^j 
\end{split}
\end{equation}
in powers of $z$.
To complete the expansion we also need to use the asymptotic expansion of $k(r)$ and $v(r)$ which is given in equation \eqref{eq:k_r_expansion}.

We obtain the following expansions for the different dimensions. For odd dimensions we only specify $h_{(d)\mu\nu}$ since it is the only one needed for the stress tensor in equation \eqref{holT} (remember that $\mathcal X$ vanishes in odd dimensions):
\begin{equation}
\begin{split}
h_{(d)\mu\nu} dx^\mu dx^\nu= &
 \left(\frac{(d-1) (x_n^{d-2}-x_n^d )}{d L^d }
\right) d\tau^2
+
 \left(
\frac{x_n^d-x_n^{d-2}}{ d L^{d}}  \right) \frac{d\rho^2}{\rho^2}
\\ &
+\frac{4 }{d} \left(   \frac{x_n^d-x_n^{d-2}+d\beta_n}{L^d (d-2)}
\right) \del_i K^b x_b \frac{d\rho}{\rho} dy^i
\\ &
+\frac{2 }{d }\left[  \left( \frac{x_n^d-x_n^{d-2}}{2 L^d } \right) \delta_{ij} + \frac{\left(x_n^d-x_n^{d-2}+d\beta_n \right)}{L^d}\tilde K^a_{ij} x_a \right]\frac{dy^i dy^j}{\rho^2}
 +\ldots.
\end{split}
\end{equation}
In $d=3$ the traceless part of the extrinsic curvature should be set to zero.
In $d=4$ we obtain
\begin{align}
\begin{split}
h_{(0)\mu\nu} dx^\mu dx^\nu= &
 d\tau^2
+
\frac{d\rho^2}{\rho^2}
+ \frac{2}{\rho} \del_i K^b x_b \, d\rho \, dy^i
+\frac{1}{\rho^2} \left[  \delta_{ij}
+ 2\tilde K^a_{ij} x_a \right] dy^i dy^j
 +\ldots,
\end{split}
\\
\begin{split}
h_{(2)\mu\nu} dx^\mu dx^\nu= &
 -\frac{1}{2 L^2}
 d\tau^2
+ \frac{1}{2L^2} \frac{d\rho^2}{\rho^2}
+ \frac{1}{2 L^2 \rho^2} \delta_{ij} dy^i dy^j
 +\ldots,
\end{split}
\\
\begin{split}
h_{(4)\mu\nu} dx^\mu dx^\nu= &
 \left(\frac{-12 x_n^4 +12 x_n^2+1}{16 L^4}
\right) d\tau^2
+
\left(\frac{1-2x_n^2}{4 L^4} \right)^2 \frac{d\rho^2}{\rho^2}
\\ &
+ \left(  \frac{16 \beta_n + \left(1-2 x_n^2\right)^2}{8 L^4}
\right) \del_i K^b x_b \frac{d\rho}{\rho} dy^i
\\ &
+\frac{1}{\rho^2}\left[  \left(\frac{1-2x_n^2}{4 L^4} \right)^2 \delta_{ij}
+ \left(  \frac{16 \beta_n + \left(1-2 x_n^2\right)^2}{8 L^4}
\right) \tilde K^a_{ij} x_a \right]dy^i dy^j
 +\ldots.
\end{split}
\end{align}
In $d=6$ instead
\begin{align}
\begin{split}
h_{(0)\mu\nu} dx^\mu dx^\nu & =  d\tau^2 +\frac{d\rho^2}{\rho^2}
+ \frac{1}{\rho} \del_i K^b x_b \, d\rho \, dy^i + \frac{1}{\rho^2} [ \delta_{ij} + 2 K_{aij}x^a ] dy^i dy^j + \cdots \,,
\end{split}
\\
\begin{split}
h_{(2)\mu\nu} dx^\mu dx^\nu & = -\frac{1}{2 L^2} d\tau^2 +\frac{1}{2 L^2} \frac{d\rho^2}{\rho^2} + \frac{1}{2 L^2 \rho^2} \delta_{ij} dy^i dy^j + \cdots \,,
\end{split}
\\
\begin{split}
h_{(4)\mu\nu} dx^\mu dx^\nu & =  \frac{1}{16 L^4} d\tau^2 +\frac{1}{16 L^4} \frac{d\rho^2}{\rho^2}
- \frac{1}{16 L^4 \rho} \del_i K^b x_b \, d\rho \, dy^i
\\
&
+ \frac{1}{16 L^4 \rho^2} [ \delta_{ij} - 2 K_{aij}x^a ] dy^i dy^j + \cdots \,,
\end{split}
 \\
 \begin{split}
h_{(6)\mu\nu} dx^\mu dx^\nu & =
 - \frac{5(x_n^6-x_n^4)}{6 L^6} d\tau^2 + \left(  \frac{x_n^6-x_n^4}{6 L^6} \right)\frac{d\rho^2}{\rho^2}  \\
 &+ \frac{1}{ L^6 \rho} \left( \beta_n +\frac{8 x_n^6-8 x_n^4+3}{48} \right) \del_i K^b x_b \, d\rho \, dy^i
 \\
& + \frac{1}{\rho^2} \left[
\left(  \frac{x_n^6-x_n^4}{6  L^6} \right) \delta_{ij} + \left( \frac{2\beta_n}{ L^6} + \frac{8 x_n^6-8 x_n^4+3}{24  L^6} \right) K_{aij}x^a \right] dy^i dy^j + \cdots \,.
\end{split}
\end{align}
Using equations \eqref{holT} and (3.15)-(3.16) of \cite{deHaro:2000vlm} we obtain the expectation value of the stress tensor of the form \eqref{Tgoodframe} where $g_n$ and $k_n$ are given by eq.~\eqref{kghol}. Note that in order to obtain the values of the anomalous contributions appearing in footnote \ref{footnote} we need to use the lower-order metrics appearing in this Appendix.

\allowdisplaybreaks

\allowdisplaybreaks




\vspace{0.5cm}


\begin{thebibliography}{10}

\bibitem{Calabrese:2004eu}
P.~Calabrese and J.~L. Cardy, {\it {Entanglement entropy and quantum field
  theory}},  {\em J. Stat. Mech.} {\bf 0406} (2004) P06002
  [\href{http://arXiv.org/abs/hep-th/0405152}{{\tt hep-th/0405152}}].

\bibitem{Calabrese:2005in}
P.~Calabrese and J.~L. Cardy, {\it {Evolution of entanglement entropy in
  one-dimensional systems}},  {\em J. Stat. Mech.} {\bf 0504} (2005) P04010
  [\href{http://arXiv.org/abs/cond-mat/0503393}{{\tt cond-mat/0503393}}].

\bibitem{Calabrese:2005zw}
P.~Calabrese and J.~L. Cardy, {\it {Entanglement entropy and quantum field
  theory: A Non-technical introduction}},  {\em Int. J. Quant. Inf.} {\bf 4}
  (2006) 429 [\href{http://arXiv.org/abs/quant-ph/0505193}{{\tt
  quant-ph/0505193}}].

\bibitem{qg1}
M.~Van~Raamsdonk, {\it {Building up spacetime with quantum entanglement}},
  {\em Gen. Rel. Grav.} {\bf 42} (2010) 2323--2329
  [\href{http://arXiv.org/abs/1005.3035}{{\tt 1005.3035}}]. [Int. J. Mod.
  Phys.D19,2429(2010)].

\bibitem{qg2}
E.~Bianchi and R.~C. Myers, {\it {On the Architecture of Spacetime Geometry}},
  {\em Class. Quant. Grav.} {\bf 31} (2014) 214002
  [\href{http://arXiv.org/abs/1212.5183}{{\tt 1212.5183}}].

\bibitem{qg3}
J.~Maldacena and L.~Susskind, {\it {Cool horizons for entangled black holes}},
  {\em Fortsch. Phys.} {\bf 61} (2013) 781--811
  [\href{http://arXiv.org/abs/1306.0533}{{\tt 1306.0533}}].

\bibitem{qg4}
T.~Jacobson, {\it {Entanglement equilibrium and the Einstein equation}},
  \href{http://arXiv.org/abs/1505.04753}{{\tt 1505.04753}}.

\bibitem{qibook}
M.~A. Nielsen and I.~L. Chuang, {\em {Quantum Computation and Quantum
  Information}}.
\newblock 2000.

\bibitem{renyi1}
A.~R\'enyi, {\it On measures of entropy and information},  in {\em Fourth
  Berkeley symposium on mathematical statistics and probability}, vol.~1,
  pp.~547--561, 1961.

\bibitem{renyi2}
A.~R{\'e}nyi, {\it On the foundations of information theory},  {\em Revue de
  l'Institut International de Statistique} (1965) 1--14.

\bibitem{Cardy:2007mb}
J.~L. Cardy, O.~A. Castro-Alvaredo and B.~Doyon, {\it {Form factors of
  branch-point twist fields in quantum integrable models and entanglement
  entropy}},  {\em J. Statist. Phys.} {\bf 130} (2008) 129--168
  [\href{http://arXiv.org/abs/0706.3384}{{\tt 0706.3384}}].

\bibitem{Hung:2014npa}
L.-Y. Hung, R.~C. Myers and M.~Smolkin, {\it {Twist operators in higher
  dimensions}},  {\em JHEP} {\bf 10} (2014) 178
  [\href{http://arXiv.org/abs/1407.6429}{{\tt 1407.6429}}].

\bibitem{roger1}
M.~B. Hastings, I.~Gonz{\'a}lez, A.~B. Kallin and R.~G. Melko, {\it Measuring
  renyi entanglement entropy in quantum monte carlo simulations},  {\em
  Physical review letters} {\bf 104} (2010), no.~15 157201.

\bibitem{roger2}
A.~B. Kallin, M.~B. Hastings, R.~G. Melko and R.~R. Singh, {\it Anomalies in
  the entanglement properties of the square-lattice heisenberg model},  {\em
  Physical Review B} {\bf 84} (2011), no.~16 165134.

\bibitem{roger3}
A.~B. Kallin, E.~Stoudenmire, P.~Fendley, R.~R. Singh and R.~G. Melko, {\it
  Corner contribution to the entanglement entropy of an o (3) quantum critical
  point in 2+ 1 dimensions},  {\em Journal of Statistical Mechanics: Theory and
  Experiment} {\bf 2014} (2014), no.~6 P06009.

\bibitem{dima}
D.~A. Abanin and E.~Demler, {\it {Measuring Entanglement Entropy of a Generic
  Many-Body System with a Quantum Switch}},  {\em Phys. Rev. Lett.} {\bf 109}
  (2012) 020504 [\href{http://arXiv.org/abs/1204.2819}{{\tt 1204.2819}}].

\bibitem{expEE}
R.~Islam, R.~Ma, P.~M. Preiss, M.~E. Tai, A.~Lukin, M.~Rispoli and M.~Greiner,
  {\it {Measuring entanglement entropy through the interference of quantum
  many-body twins}},  \href{http://arXiv.org/abs/1509.01160}{{\tt 1509.01160}}.

\bibitem{Ryu:2006bv}
S.~Ryu and T.~Takayanagi, {\it {Holographic derivation of entanglement entropy
  from AdS/CFT}},  {\em Phys.Rev.Lett.} {\bf 96} (2006) 181602
  [\href{http://arXiv.org/abs/hep-th/0603001}{{\tt hep-th/0603001}}].

\bibitem{Ryu:2006ef}
S.~Ryu and T.~Takayanagi, {\it {Aspects of Holographic Entanglement Entropy}},
  {\em JHEP} {\bf 0608} (2006) 045
  [\href{http://arXiv.org/abs/hep-th/0605073}{{\tt hep-th/0605073}}].

\bibitem{Hubeny:2007xt}
V.~E. Hubeny, M.~Rangamani and T.~Takayanagi, {\it {A Covariant holographic
  entanglement entropy proposal}},  {\em JHEP} {\bf 0707} (2007) 062
  [\href{http://arXiv.org/abs/0705.0016}{{\tt 0705.0016}}].

\bibitem{Hung:2011xb}
L.-Y. Hung, R.~C. Myers and M.~Smolkin, {\it {On Holographic Entanglement
  Entropy and Higher Curvature Gravity}},  {\em JHEP} {\bf 1104} (2011) 025
  [\href{http://arXiv.org/abs/1101.5813}{{\tt 1101.5813}}].

\bibitem{deBoer:2011wk}
J.~de~Boer, M.~Kulaxizi and A.~Parnachev, {\it {Holographic Entanglement
  Entropy in Lovelock Gravities}},  {\em JHEP} {\bf 07} (2011) 109
  [\href{http://arXiv.org/abs/1101.5781}{{\tt 1101.5781}}].

\bibitem{Dong:2013qoa}
X.~Dong, {\it {Holographic Entanglement Entropy for General Higher Derivative
  Gravity}},  {\em JHEP} {\bf 1401} (2014) 044
  [\href{http://arXiv.org/abs/1310.5713}{{\tt 1310.5713}}].

\bibitem{Camps:2013zua}
J.~Camps, {\it {Generalized entropy and higher derivative Gravity}},  {\em
  JHEP} {\bf 1403} (2014) 070 [\href{http://arXiv.org/abs/1310.6659}{{\tt
  1310.6659}}].

\bibitem{Faulkner:2013ana}
T.~Faulkner, A.~Lewkowycz and J.~Maldacena, {\it {Quantum corrections to
  holographic entanglement entropy}},  {\em JHEP} {\bf 11} (2013) 074
  [\href{http://arXiv.org/abs/1307.2892}{{\tt 1307.2892}}].

\bibitem{Lewkowycz:2013nqa}
A.~Lewkowycz and J.~Maldacena, {\it {Generalized gravitational entropy}},  {\em
  JHEP} {\bf 1308} (2013) 090 [\href{http://arXiv.org/abs/1304.4926}{{\tt
  1304.4926}}].

\bibitem{DLR}
X.~Dong, A.~Lewkowycz and M.~Rangamani, {\it {Deriving covariant holographic
  entanglement}},  {\em to appear}.

\bibitem{Dong:2016fnf}
X.~Dong, {\it {An Area-Law Prescription for Holographic Renyi Entropies}},
  \href{http://arXiv.org/abs/1601.06788}{{\tt 1601.06788}}.

\bibitem{Casini:2011kv}
H.~Casini, M.~Huerta and R.~C. Myers, {\it {Towards a derivation of holographic
  entanglement entropy}},  {\em JHEP} {\bf 1105} (2011) 036
  [\href{http://arXiv.org/abs/1102.0440}{{\tt 1102.0440}}].

\bibitem{Hung:2011nu}
L.-Y. Hung, R.~C. Myers, M.~Smolkin and A.~Yale, {\it {Holographic Calculations
  of Renyi Entropy}},  {\em JHEP} {\bf 12} (2011) 047
  [\href{http://arXiv.org/abs/1110.1084}{{\tt 1110.1084}}].

\bibitem{Dong:2016wcf}
X.~Dong, {\it {Shape Dependence of Holographic R\'enyi Entropy in Conformal
  Field Theories}},  {\em Phys. Rev. Lett.} {\bf 116} (2016), no.~25 251602
  [\href{http://arXiv.org/abs/1602.08493}{{\tt 1602.08493}}].

\bibitem{Camps:2016gfs}
J.~Camps, {\it {Gravity duals of boundary cones}},
  \href{http://arXiv.org/abs/1605.08588}{{\tt 1605.08588}}.

\bibitem{Bianchi:2015liz}
L.~Bianchi, M.~Meineri, R.~C. Myers and M.~Smolkin, {\it {R\'enyi entropy and
  conformal defects}},  \href{http://arXiv.org/abs/1511.06713}{{\tt
  1511.06713}}.

\bibitem{mark0}
M.~Mezei, {\it {Entanglement entropy across a deformed sphere}},  {\em Phys.
  Rev.} {\bf D91} (2015), no.~4 045038
  [\href{http://arXiv.org/abs/1411.7011}{{\tt 1411.7011}}].

\bibitem{Bueno:2015rda}
P.~Bueno, R.~C. Myers and W.~Witczak-Krempa, {\it {Universality of corner
  entanglement in conformal field theories}},  {\em Phys. Rev. Lett.} {\bf 115}
  (2015), no.~2 021602 [\href{http://arXiv.org/abs/1505.04804}{{\tt
  1505.04804}}].

\bibitem{Bueno:2015lza}
P.~Bueno and R.~C. Myers, {\it {Universal entanglement for higher dimensional
  cones}},  {\em JHEP} {\bf 12} (2015) 168
  [\href{http://arXiv.org/abs/1508.00587}{{\tt 1508.00587}}].

\bibitem{Carmi:2015dla}
D.~Carmi, {\it {On the Shape Dependence of Entanglement Entropy}},  {\em JHEP}
  {\bf 12} (2015) 043 [\href{http://arXiv.org/abs/1506.07528}{{\tt
  1506.07528}}].

\bibitem{Fonda:2014cca}
P.~Fonda, L.~Giomi, A.~Salvio and E.~Tonni, {\it {On shape dependence of
  holographic mutual information in AdS$_{4}$}},  {\em JHEP} {\bf 02} (2015)
  005 [\href{http://arXiv.org/abs/1411.3608}{{\tt 1411.3608}}].

\bibitem{Fonda:2015nma}
P.~Fonda, D.~Seminara and E.~Tonni, {\it {On shape dependence of holographic
  entanglement entropy in AdS$_{4}$/CFT$_{3}$}},  {\em JHEP} {\bf 12} (2015)
  037 [\href{http://arXiv.org/abs/1510.03664}{{\tt 1510.03664}}].

\bibitem{Cardy:1984bb}
J.~L. Cardy, {\it {Conformal Invariance and Surface Critical Behavior}},  {\em
  Nucl.Phys.} {\bf B240} (1984) 514--532.

\bibitem{McAvity:1995zd}
D.~McAvity and H.~Osborn, {\it {Conformal field theories near a boundary in
  general dimensions}},  {\em Nucl.Phys.} {\bf B455} (1995) 522--576
  [\href{http://arXiv.org/abs/cond-mat/9505127}{{\tt cond-mat/9505127}}].

\bibitem{Billo:2016cpy}
M.~Billò, V.~Gonçalves, E.~Lauria and M.~Meineri, {\it {Defects in conformal
  field theory}},  {\em JHEP} {\bf 04} (2016) 091
  [\href{http://arXiv.org/abs/1601.02883}{{\tt 1601.02883}}].

\bibitem{safdi1}
J.~Lee, L.~McGough and B.~R. Safdi, {\it {R\'enyi entropy and geometry}},  {\em
  Phys. Rev.} {\bf D89} (2014), no.~12 125016
  [\href{http://arXiv.org/abs/1403.1580}{{\tt 1403.1580}}].

\bibitem{Lewkowycz:2014jia}
A.~Lewkowycz and E.~Perlmutter, {\it {Universality in the geometric dependence
  of R\'enyi entropy}},  {\em JHEP} {\bf 01} (2015) 080
  [\href{http://arXiv.org/abs/1407.8171}{{\tt 1407.8171}}].

\bibitem{Bueno:2015qya}
P.~Bueno, R.~C. Myers and W.~Witczak-Krempa, {\it {Universal corner
  entanglement from twist operators}},  {\em JHEP} {\bf 09} (2015) 091
  [\href{http://arXiv.org/abs/1507.06997}{{\tt 1507.06997}}].

\bibitem{Dowker1}
J.~S. Dowker, {\it {Conformal weights of charged R\'enyi entropy twist
  operators for free scalar fields in arbitrary dimensions}},  {\em J. Phys.}
  {\bf A49} (2016), no.~14 145401 [\href{http://arXiv.org/abs/1508.02949}{{\tt
  1508.02949}}].

\bibitem{Dowker2}
J.~S. Dowker, {\it {Conformal weights of charged Renyi entropy twist operators
  for free Dirac fields in arbitrary dimensions}},
  \href{http://arXiv.org/abs/1510.08378}{{\tt 1510.08378}}.

\bibitem{Faulkner:2015csl}
T.~Faulkner, R.~G. Leigh and O.~Parrikar, {\it {Shape Dependence of
  Entanglement Entropy in Conformal Field Theories}},  {\em JHEP} {\bf 04}
  (2016) 088 [\href{http://arXiv.org/abs/1511.05179}{{\tt 1511.05179}}].

\bibitem{again}
S.~Balakrishnan, S.~Dutta and T.~Faulkner, {\it {Gravitational dual of the
  R\'{e}nyi twist displacement operator}},
  \href{http://arXiv.org/abs/1607.06155}{{\tt 1607.06155}}.

\bibitem{brandon}
R.~C. Myers and B.~Robinson, {\it {Black Holes in Quasi-topological Gravity}},
  {\em JHEP} {\bf 08} (2010) 067 [\href{http://arXiv.org/abs/1003.5357}{{\tt
  1003.5357}}].

\bibitem{deHaro:2000vlm}
S.~de~Haro, S.~N. Solodukhin and K.~Skenderis, {\it {Holographic reconstruction
  of space-time and renormalization in the AdS / CFT correspondence}},  {\em
  Commun. Math. Phys.} {\bf 217} (2001) 595--622
  [\href{http://arXiv.org/abs/hep-th/0002230}{{\tt hep-th/0002230}}].

\bibitem{Fefferman:2007rka}
C.~Fefferman and C.~R. Graham, {\it {The ambient metric}},
  \href{http://arXiv.org/abs/0710.0919}{{\tt 0710.0919}}.

\bibitem{Osborn:1993cr}
H.~Osborn and A.~Petkou, {\it {Implications of conformal invariance in field
  theories for general dimensions}},  {\em Annals Phys.} {\bf 231} (1994)
  311--362 [\href{http://arXiv.org/abs/hep-th/9307010}{{\tt hep-th/9307010}}].

\bibitem{Erdmenger:1996yc}
J.~Erdmenger and H.~Osborn, {\it {Conserved currents and the energy momentum
  tensor in conformally invariant theories for general dimensions}},  {\em
  Nucl. Phys.} {\bf B483} (1997) 431--474
  [\href{http://arXiv.org/abs/hep-th/9605009}{{\tt hep-th/9605009}}].

\bibitem{Buchel:2009sk}
A.~Buchel, J.~Escobedo, R.~C. Myers, M.~F. Paulos, A.~Sinha and M.~Smolkin,
  {\it {Holographic GB gravity in arbitrary dimensions}},  {\em JHEP} {\bf 03}
  (2010) 111 [\href{http://arXiv.org/abs/0911.4257}{{\tt 0911.4257}}].

\bibitem{wok1}
M.~Brigante, H.~Liu, R.~C. Myers, S.~Shenker and S.~Yaida, {\it {The Viscosity
  Bound and Causality Violation}},  {\em Phys. Rev. Lett.} {\bf 100} (2008)
  191601 [\href{http://arXiv.org/abs/0802.3318}{{\tt 0802.3318}}].

\bibitem{wok2}
A.~Buchel and R.~C. Myers, {\it {Causality of Holographic Hydrodynamics}},
  {\em JHEP} {\bf 08} (2009) 016 [\href{http://arXiv.org/abs/0906.2922}{{\tt
  0906.2922}}].

\bibitem{wok3}
D.~M. Hofman, {\it {Higher Derivative Gravity, Causality and Positivity of
  Energy in a UV complete QFT}},  {\em Nucl. Phys.} {\bf B823} (2009) 174--194
  [\href{http://arXiv.org/abs/0907.1625}{{\tt 0907.1625}}].

\bibitem{Camanho:2014apa}
X.~O. Camanho, J.~D. Edelstein, J.~Maldacena and A.~Zhiboedov, {\it {Causality
  Constraints on Corrections to the Graviton Three-Point Coupling}},  {\em
  JHEP} {\bf 02} (2016) 020 [\href{http://arXiv.org/abs/1407.5597}{{\tt
  1407.5597}}].

\bibitem{Myers:2010xs}
R.~C. Myers and A.~Sinha, {\it {Seeing a c-theorem with holography}},  {\em
  Phys. Rev.} {\bf D82} (2010) 046006
  [\href{http://arXiv.org/abs/1006.1263}{{\tt 1006.1263}}].

\bibitem{Myers:2010tj}
R.~C. Myers and A.~Sinha, {\it {Holographic c-theorems in arbitrary
  dimensions}},  {\em JHEP} {\bf 01} (2011) 125
  [\href{http://arXiv.org/abs/1011.5819}{{\tt 1011.5819}}].

\bibitem{Faulkner:2013ica}
T.~Faulkner, M.~Guica, T.~Hartman, R.~C. Myers and M.~Van~Raamsdonk, {\it
  {Gravitation from Entanglement in Holographic CFTs}},  {\em JHEP} {\bf 03}
  (2014) 051 [\href{http://arXiv.org/abs/1312.7856}{{\tt 1312.7856}}].

\bibitem{Lewkowycz:2013laa}
A.~Lewkowycz and J.~Maldacena, {\it {Exact results for the entanglement entropy
  and the energy radiated by a quark}},  {\em JHEP} {\bf 1405} (2014) 025
  [\href{http://arXiv.org/abs/1312.5682}{{\tt 1312.5682}}].

\bibitem{Fiol:2015spa}
B.~Fiol, E.~Gerchkovitz and Z.~Komargodski, {\it {The Exact Bremsstrahlung
  Function in N=2 Superconformal Field Theories}},
  \href{http://arXiv.org/abs/1510.01332}{{\tt 1510.01332}}.

\bibitem{Maldacena:2015waa}
J.~Maldacena, S.~H. Shenker and D.~Stanford, {\it {A bound on chaos}},
  \href{http://arXiv.org/abs/1503.01409}{{\tt 1503.01409}}.

\bibitem{Chu:2016tps}
C.-S. Chu and R.-X. Miao, {\it {Universality in the Shape Dependence of
  Holographic R\'enyi Entropy for General Higher Derivative Gravity}},
  \href{http://arXiv.org/abs/1608.00328}{{\tt 1608.00328}}.

\bibitem{Lee:2014zaa}
J.~Lee, A.~Lewkowycz, E.~Perlmutter and B.~R. Safdi, {\it {R\'enyi entropy,
  stationarity, and entanglement of the conformal scalar}},  {\em JHEP} {\bf
  03} (2015) 075 [\href{http://arXiv.org/abs/1407.7816}{{\tt 1407.7816}}].

\bibitem{Hofman:2008ar}
D.~M. Hofman and J.~Maldacena, {\it {Conformal collider physics: Energy and
  charge correlations}},  {\em JHEP} {\bf 05} (2008) 012
  [\href{http://arXiv.org/abs/0803.1467}{{\tt 0803.1467}}].

\bibitem{future}
L.~Bianchi~{\it et al}, {\it {On the Shape Dependence of R\'enyi Entropy in
  Free CFTs}},  {\em in preparation}.

\bibitem{max}
M.~A. Metlitski, C.~A. Fuertes and S.~Sachdev, {\it {Entanglement Entropy in
  the O(N) model}},  {\em Phys. Rev.} {\bf B80} (2009), no.~11 115122
  [\href{http://arXiv.org/abs/0904.4477}{{\tt 0904.4477}}].

\bibitem{Belin:2013dva}
A.~Belin, A.~Maloney and S.~Matsuura, {\it {Holographic Phases of Renyi
  Entropies}},  {\em JHEP} {\bf 12} (2013) 050
  [\href{http://arXiv.org/abs/1306.2640}{{\tt 1306.2640}}].

\bibitem{pufu}
S.~S. Pufu. Private communication.

\end{thebibliography}
\providecommand{\href}[2]{#2}\begingroup\raggedright\endgroup

\end{document}